\documentclass{article}

\usepackage[english]{babel}

\usepackage[letterpaper,top=2cm,bottom=2cm,left=2.5cm,right=2.5cm,marginparwidth=1.75cm]{geometry}

\usepackage{amsmath, amsfonts, amssymb, amsthm}
\usepackage[normalem]{ulem}
\usepackage{graphicx}
\usepackage{bbold}
\usepackage{stmaryrd}
\usepackage[colorlinks=true, allcolors=blue]{hyperref}
\usepackage{xspace}
\usepackage[color=yellow]{todonotes}
\usepackage[section]{placeins}  
\usepackage{authblk}

\newtheorem{remark}{Remark}[section]

\theoremstyle{definition}

\newcommand{\ito}{It\^o\xspace}

\newcommand{\E}{\mathbb{E}}

\newcommand{\MM}{\mathcal{M}}
\newcommand{\MW}{\mathcal{W}}

\newcommand{\MV}{\mathcal{V}}

\newcommand{\MT}{\mathcal{T}}
\newcommand{\MP}{\mathcal{P}}

\newcommand{\bu}{\mathbf{u}}
\newcommand{\bn}{\mathbf{n}}

\newcommand{\BD}{\mathbf{D}}
\newcommand{\bx}{\mathbf{x}}
\newcommand{\bxi}{\boldsymbol{\xi}}
\newcommand{\td}{\text{d}}
\newcommand{\gradperp}{\nabla^\perp}
\newcommand{\grad}{\nabla}

\newcommand{\sdiff}{\mathrm{SDiff}}

\newcommand{\dw}{\text{d}W_t^i}
\newcommand{\dt}{\td t}

\newcommand{\Nabla}{\nabla}

\newcommand{\ad}{\mathrm{ad}}

\newcommand{\EM}{\mathrm{EM}}
\newcommand{\DW}{\Delta W_n^i}
\newcommand{\bk}{\mathbf{k}}
\newcommand{\bm}{\mathbf{m}}
\newcommand{\atan}{\mathrm{atan2}}

\graphicspath{{./figures_salt_sflt_comparison/}}

\title{Numerical comparison of energy- versus circulation-preserving stochastic vortex dynamics}

\author[1]{Sagy Ephrati\thanks{Corresponding author. Email: \texttt{s.ephrati@imperial.ac.uk}}}
\author[1]{Darryl D. Holm\thanks{Email: \texttt{d.holm@imperial.ac.uk}}}
\affil[1]{Department of Mathematics, Imperial College London, London SW7 2AZ, UK}
\date{}

\begin{document}
\maketitle

\begin{abstract}
    We compare two geometric stochastic frameworks for the two-dimensional Euler equations, being the circulation-preserving stochastic advection by Lie transport (SALT) and the energy-preserving stochastic forcing by Lie transport (SFLT) approaches.
    While preserving both circulation and energy is ideal, their simultaneous conservation restricts perturbations to a stochastic reparametrization of time.
    Consequently, a fundamental choice must be made between preserving structure or the kinetic energy.
    Analysis reveals that SALT is significantly more sensitive to high-frequency flow components, with noise effects scaling by $| \bk |^2$ relative to SFLT.
    This suggests that SALT acts as a localized perturbation sensitive to sharp gradients, while SFLT behaves as a more regularized global forcing.
    Numerical experiments on a traveling dipole, vortex merger, and forced-damped turbulence confirm that SALT introduces uncertainty localized near dynamically active vorticity gradients, whereas SFLT produces a more diffuse variance field spread across the domain.
    These results illustrate how the choice of geometric invariant fundamentally determines scale-sensitivity and spatial distribution of modeled uncertainty in vortex dynamics.
\end{abstract}

\section{Introduction}

The accurate prediction of multi-scale dynamics in fluid flows is a major computational challenge. Unresolvable scales of motion necessitate closure models that account for the influence of sub-grid processes on the resolved flow components. In practice, the inability to resolve all scales of motion results in the loss of fine-scale information. However, some aspects of uncertainty may be represented through the use of stochastic models. This viewpoint has long been recognized in numerical weather prediction and climate modeling \cite{hasselmann1976stochastic}, and is increasingly gaining traction in turbulence modeling and large-eddy simulation \cite{freitas2026importance, ephrati2025probabilistic}.
Simultaneously, the long-time accuracy and physical fidelity of geometric methods have caused increased interest in the numerical preservation of physical conservation laws.

The introduction of stochasticity should not come at the cost of physical consistency. Hence, recent emphasis on geometric methods has encouraged the development of geometric stochastic dynamics in which key invariants of the underlying deterministic system are preserved. A compelling approach particularly suited for ideal fluid dynamics is to use \textit{transport noise} \cite{holm2015variational}, where stochasticity is added to the advection velocity rather than as an additive force. 
While structure-preserving stochasticity offers advantages, it generally necessitates a trade-off between energy and geometric structure, since any approach that preserves both simultaneously is effectively a stochastic reparametrization of time \cite{zhong1988lie}. In this work, we adopt two canonical approaches to energy- and structure-preservation in the two-dimensional Euler equations, comparing the influence of these different parameterizations on vortex dynamics. Specifically, we investigate Stochastic Advection by Lie Transport (SALT) \cite{holm2015variational}, which preserves the underlying Lie--Poisson structure and its associated Casimirs, and also Stochastic Forcing by Lie Transport (SFLT) \cite{holm2021stochastic}, which is formulated to ensure exact preservation of kinetic energy. 
To bridge the gap between these stochastic frameworks and practical implementation in multi-scale modeling, we also consider their mean-field approximations, Lagrangian-Averaged SALT (LA SALT) \cite{drivas2019lagrangian, alonso2020modelling} and Eulerian-Averaged SFLT (EA SFLT) \cite{hu2021variational}, which regularize the dynamics through non-local probabilistic averaging.

Recent years have seen an increase in the understanding of the diffusive properties of transport noise. Tracing back to stochastic passive scalar advection \cite{kraichnan1994anomalous}, more recent analytical works have proven convergence of stochastic transport equations to parabolic equations \cite{galeati2020convergence}. Additionally, defining a stochastic transport term as the superposition of high-frequency Fourier modes with carefully selected magnitudes has been proven to converge to standard diffusion in two-dimensional turbulence \cite{flandoli2021scaling}. The diffusive properties of transport noise have recently also been demonstrated numerically \cite{cifani2025diffusion, ephrati2026diffusive, cifani2026anomalous, flandoli2022effect}.

There have been several numerical investigations into SALT in the context of sub-grid scale modeling, uncertainty quantification, and data assimilation for geophysical fluid-dynamical problems.
A multi-scale data-driven procedure for the calibration of SALT terms for the two-dimensional Euler equations was presented by \cite{cotter2019numerically}, using POD modes and time-correlated processes \cite{ephrati2023noise} to quantify uncertainty induced by coarsening.
SALT was combined with a particle filter in quasi-geostrophic flow to significantly reduce the computational degrees of freedom without sacrificing accuracy \cite{cotter2018modelling}.
The SALT framework has also been applied to the thermal quasi-geostrophic equations \cite{holm2024comparing} to compare kinetic energy coupling with stochastic potential energy coupling (SPEC), both of which maintain Casimir conservation. This revealed that SALT yields a significantly higher buoyancy variance, though this discrepancy diminishes over longer time scales as buoyancy and bathymetry interact. More recently, Sharma and Korn \cite{sharma2026structure} introduced SALT into a coupled ocean atmosphere model and employed data-calibrated scalar Ornstein--Uhlenbeck processes in their stochastic forcing, and found a good spread-error relationship in their ensemble forecasts.

To date, the assessment of SFLT in numerical studies has been limited. 
A first study was carried out by \cite{hu2021variational}, who compared the effects of SALT and SFLT on the primitive equations for oceanic dynamics. A low-rank stochastic model was calibrated from Lagrangian and Eulerian differences between the true velocity and filtered velocity, and large sensitivity with respect to the calibration data was reported.
The work by \cite{cifani2022sparse} investigated model reduction of the two-dimensional Euler equations on the sphere, employing stochastic SALT and SFLT as stochastic parameterizations to close the equations of motion for large-scale dynamics. Notably, the term parametrized by SFLT was observed to be larger than that parametrized by SALT. This, along with a nonzero energy exchange between resolved and modeled scales, caused the SFLT parameterization to deviate from the reference results.
Recently, the SALT and SFLT frameworks have been compared numerically for the Camassa--Holm equation \cite{holm2026comparative}. Characteristic solution behavior of this system is the emergence of localized solitary wave-like structures referred to as `peakons'. By employing Fourier modes multiplied by independent stochastic increments in the stochastic forcing, the SFLT approach generated an ensemble of solutions spread around the deterministic solution. Simultaneously, SALT induced a change in the magnitudes and speeds of the peakons, resulting in markedly different solutions.

\paragraph*{Objectives}
While SALT has matured through numerical investigations in geophysical contexts, the assessment of SFLT remains relatively sparse. A comprehensive comparison of the physics induced by these structure-preserving and energy-preserving frameworks in two-dimensional flows is currently missing. This work seeks to fill that gap, providing a foundation to better understand and more easily interpret noise-induced phenomena in complex flows, such as thermal convection \cite{holm2023deterministic} and magnetohydrodynamics \cite{holm2024deterministic}, in future work. 

The primary objective of this work is to provide a comparison between energy-preserving and structure-preserving stochastic parameterizations for the two-dimensional incompressible Euler equations. Specifically, we evaluate the performance of (LA) SALT and (EA) SFLT across several test cases of vortex interactions to discern their respective impacts on uncertainty quantification. We focus particularly on the physics induced by these parameterizations. By studying these interactions, both through analysis and numerical simulation, we aim to gain insights into how different geometric forcings affect flow patterns. These findings may inform future modeling choices and data-based calibration methodologies.

\paragraph*{Structure}
The paper is structured as follows. 
The deterministic two-dimensional Euler equations are introduced in Section \ref{sec:deterministic_equations}, and their geometry and relation to geophysical fluid dynamics and plasma physics are summarized.
In Section \ref{sec:stochastic_equations}, we introduce the stochastic methods used in this paper and provide two theoretical perspectives for comparison.
The numerical tests are described in Section \ref{sec:deterministic_equations} and the paper is concluded in Section \ref{sec:conclusions}.

\section{Deterministic Euler equations} \label{sec:deterministic_equations}
The incompressible two-dimensional Euler equations are central to this work. We briefly outline the governing equations, their application in geophysical fluid dynamics and plasma physics, and their geometric properties.

\paragraph{Governing equations}
In the stream function-vorticity formulation, these equations read \begin{align}
    \partial_t\omega &= \{\psi, \omega\} 
    \label{eq:vorticity_advection}\\
    \Delta\psi &= -\omega, \label{eq:strm_vort_relation}
\end{align}
where $\psi$ is the stream function and $\omega$ is the vorticity. These quantities are related to the velocity $\bu$ via $\gradperp\psi=\bu$ and $\omega=\gradperp\cdot\bu$, where $\gradperp=(-\partial_y, \partial_x)^T$ is the perpendicular gradient. 

We define the equations on a flat two-dimensional domain, employing two types of boundary conditions depending on the considered test case. The first type is a periodic boundary condition in both directions, the second is an impenetrable free-slip condition for the velocity, defined as \begin{equation}
    \psi\big|_{\partial\mathcal{M}}=0. \label{eq:bc}
\end{equation}

The advection term in \eqref{eq:vorticity_advection} is written using the Poisson bracket. Given two functions $f, g:\mathcal{M}\to\mathbb{R}$, the bracket is defined as \begin{equation}
    \{f, g\} = \gradperp f\cdot \nabla g. \label{eq:poisson_bracket}
\end{equation}
The Poisson bracket is skew-symmetric and satisfies the Jacobi identity. As will be highlighted below, these properties are fundamental to conservation of energy and an infinite family of Casimir functions under the deterministic flow.

\paragraph{The Euler equations in GFD and plasma physics}
The two-dimensional Euler equations arise in the inviscid limit of the two-dimensional Navier--Stokes equations and serve as a building block for dynamical systems in geophysical fluid dynamics and plasma physics.
In geophysical fluid dynamics, the two-dimensional Euler equations can be obtained via a sequence of simplifying assumptions when starting from the three-dimensional Euler for stratified, rotating fluids \cite{holm2021stochastic}. By assuming small buoyancy stratification and hydrostatic vertical pressure, one respectively obtains the Boussinesq equations and the primitive equations. Both models are defined in three spatial dimensions and can be reduced to the Green--Naghdi equations and the shallow water equations by averaging in the vertical direction and assuming variations in the free surface height. Subsequently, the quasi-geostrophic equations are obtained via asymptotic expansion around geostrophic balance.

Despite these different levels of complexity, the Euler equations retain the essential nonlinear advection present in two-dimensional GFD and MHD models. Specifically, the dynamics of coherent vortices in the Euler equations can be traced back to the advection of (potential) vorticity, and is therefore closely mirrored by eddies in the QG equations and traveling dipole solutions (modons) in the (thermal) shallow water equations \cite{lahaye2024equatorial} and shallow water magnetohydrodynamics \cite{lahaye2022coherent}.
Furthermore, the energy and enstrophy are conserved in two-dimensional Euler dynamics, which gives rise to the characteristic dual cascade of two-dimensional turbulence. The simultaneous movement of energy to small wave numbers and enstrophy to large wave numbers causes self-organization observed in planetary-scale flows \cite{zeitlin2018geophysical}.
This dual cascade mechanism was famously explored from a statistical physics perspective by Kraichnan \cite{kraichnan1967inertial}, who showed that the conservation properties constrain the triadic interactions and lead to the emergence of large-scale coherent structures in two-dimensional turbulent flow.
\cite{kraichnan1967inertial}

\paragraph{Geometry of the two-dimensional Euler equations}
In the absence of external forcing and damping, the equations (\ref{eq:vorticity_advection}-\ref{eq:strm_vort_relation}) conserve the kinetic energy $E$ and an infinite family of Casimir functions $C_n$, \begin{align}
    E(\omega) &= \frac{1}{2}\int_\Omega\! |\gradperp\psi|^2\,\td\Omega = -\frac{1}{2}\int_\Omega\, \psi\omega\,\td\Omega, \label{eq:energy}\\
    C_n(\omega) &= \frac{1}{2}\int_\Omega\! \omega^n\,\td\Omega.
\end{align} 
A well-known Casimir function is the enstrophy $C_2$. These conserved quantities arise naturally from the geometric interpretation of hydrodynamics. We summarize this perspective below; for full details we refer to \cite{arnold1998topological}.

The cornerstone of geometric hydrodynamics is Arnold's demonstration \cite{arnold1966geometrie} that the Euler equations exactly describe geodesic motion on the group $\sdiff(\MM)$ of volume-preserving diffeomorphisms, with respect to the kinetic energy metric. Here, $\MM$ is a Riemannian manifold to which the fluid movement is confined.
The essential step is to treat $\sdiff(\MM)$ as an infinite-dimensional Lie group whose action is defined as composition of functions\footnote{Composition will be denoted by a $\circ$ symbol. The same symbol is used in stochastic (partial) differential equations to denote Stratonovich integration. The meaning of the symbol will be clear from the context.} as the group action.
Then, for each diffeomorphism $\Phi\in\sdiff(\MM)$, elements of the tangent space $T_\Phi\MM$ are given by $\bu\circ\Phi$, where $\bu$ is a divergence-free vector field on $\MM$. A metric for $T_\Phi\MM$ is then found by defining \begin{equation}
    \langle \bu\circ\Phi,\mathbf{v}\circ\Phi\rangle_\Phi = \int_M\!\bu\cdot\mathbf{v}, \label{eq:metric_sdiff}
\end{equation}
i.e., a right invariant $L^2$ inner product. This defines a smoothly varying inner product on $\MM$ and therefore (by definition) a Riemannian metric, which provides a notion of distance for $\sdiff(\MM)$. The key insight of Arnold was that a solution $\bu(t)$ to the Euler equations is a geodesic for the metric \eqref{eq:metric_sdiff}, meaning that trajectories traced out by $\bu(t)$ locally minimize length in $\sdiff(\MM)$.

Alternatively, changing the metric yields distinct geodesic solutions described by modified dynamics. For example, replacing the $L^2$ inner product by an $H^1$ inner product yields regularized dynamics via a nonlinear dispersive modification \cite{holm1998euler}. In two-dimensional turbulence, this approach preserves the conserved quantities and allows for precise control over the energy levels in resolved scales of motion \cite{ephrati2025spectral}. That is, these regularized solutions remain on the coadjoint orbits, the solution set with constant Casimirs shared by the `standard' two-dimensional Euler equations, while tracing out an alternative trajectory due to the change of the metric. In the governing equations, this is reflected by a change in the advecting velocity. As we shall see in the next section, the advecting velocity may instead also be perturbed stochastically, resulting in stochastic dynamics with conserved quantities.

The Euler equations can also be viewed as a \textit{Lie--Poisson} system on the dual of the Lie algebra \cite{marsden1999introduction}. Let $G$ denote a Lie group with $\mathfrak{g}=T_eG$ the corresponding Lie algebra, and assume a right-invariant metric on at $g\in G$ denoted by $\langle\cdot,\cdot\rangle_g$. The length of a tangent vector $\dot g = vg\in T_gG$, $g\in G, v\in \mathfrak{g}$ is then expressed as $\langle \dot g, \dot g\rangle_g=\langle \dot g g^{-1}, \dot g g^{-1}\rangle_e = \langle v, v\rangle_e$. With this identity in place, the momentum variable $\mu=\langle v,\cdot\rangle$ satisfies the equation \begin{equation}
    \dot\mu + \ad_v^*\mu = 0.
\end{equation}
Here, $\ad_v^*:\mathfrak{g}^*\to\mathfrak{g}^*$ is the infinitesimal action of the Lie algebra on its dual. For the Lie algebra of smooth, zero-mean functions on the torus equipped with the Poisson bracket \eqref{eq:poisson_bracket} and inner product defined by \begin{equation}
    \langle v, w\rangle_e = \int_\Omega \! \grad v\cdot\grad w\,\td\Omega,
\end{equation}
the momentum variable is defined as $\omega=\langle\psi,\cdot\rangle_e$ and $\ad_\psi^* \omega = \{\omega,\psi\}$ \cite{modin2025brief, arnold1998topological}. Defining the Hamiltonian as the kinetic energy \eqref{eq:energy}, i.e., $H(\omega)=E(\omega)$, the Lie--Poisson formulation of the two-dimensional Euler equations reads \begin{equation}
    \dot\omega + \ad_{\frac{\delta H}{\delta \omega}}\omega = 0.
\end{equation}
A change of metric yields a different Hamiltonian and thus changes $\delta H/\delta\omega$. However, this leaves the Lie--Poisson structure unchanged. The preservation of these crucial properties in the deterministic case provides a natural foundation for the introduction of stochasticity, which we explore in the following section.

\section{Geometric stochastic frameworks} \label{sec:stochastic_equations}
Two different kinds of stochastic forcings will be introduced into the Euler equations, following the principles of Stochastic Advection by Lie Transport (SALT) \cite{holm2015variational} and Stochastic Forcing by Lie Transport (SFLT) \cite{holm2021stochastic}. Additionally, we study the Lagrangian-averaged variant of SALT (LA SALT) \cite{alonso2020modelling}.
We introduce the frameworks and highlight their key properties below.

\paragraph{SALT} In the SALT framework, stochastic partial differential equations (SPDEs) are derived from a stochastically constrained variational principle. Here, stochasticity enters the action integral through the Lie derivative of an advected quantity. Consequently, advected quantities evolve according to a stochastically perturbed vector field. For the two-dimensional Euler equations, we replace the deterministic infinitesimal advecting velocity field $\bu\,\dt$ by a stochastically perturbed field $\bu\,\dt + \sum\bxi_i\circ\dw$. Here, $W_t^i$ denote independent Wiener processes and the symbol $\circ$ indicates that the stochastic integral is understood in the Stratonovich sense.\footnote{The Stratonovich integral is used here since the chain rule holds for Stratonovich processes. This property is allows for manifold-valued curves in differential geometry to be extended  manifold-valued processes \cite{kloeden1977numerical, emery2006two}.} We consider $\bxi_i=\bxi_i(x)$ as (fixed, space-dependent) incompressible velocity fields.

In total, the SALT--Euler equations  read \begin{equation}
    \td\omega + (\bu\,\dt + \sum_i\bxi_i\circ\dw)\cdot\nabla\omega = 0.
\end{equation}
Since $\bxi_i$ are assumed to be divergence-free, we may express these vector fields via potential functions $\zeta_i$ as $\gradperp\zeta_i=\bxi_i$ and obtain the equivalent stream function-vorticity formulation of the SALT--Euler equations, \begin{equation}
    \td\omega + \{\psi\,\dt + \sum_i\zeta_i\circ\dw, \omega\} = 0 \label{eq:SALT}
\end{equation}

This stochastic extension of the variational derivation of ideal fluid dynamics ensure that the resulting SALT-fluid equations share certain geometric properties with their deterministic counterparts. In particular, the Kelvin circulation theorem for the incompressible flows with SALT immediately reveals that circulation is conserved for loops moving with the stochastically perturbed fluid trajectories \cite{holm2015variational}. 

Equation \eqref{eq:SALT} can be cast in the Lie--Poisson formulation as \begin{equation}
    \td\omega + \ad_{\frac{\delta H}{\delta \omega}\dt + \sum_i\frac{\delta H_i}{\delta_\omega}\circ\dw}^*\omega = 0,
\end{equation}
where $H_i$ are noise Hamiltonians satisfying $\delta H_i/\delta\omega = \zeta_i$.
This shows that SALT preserves the Lie--Poisson structure of the deterministic equations and as a consequence, the Casimir functions $C_n(\omega)=\int_\Omega\omega^n\,\td\Omega$ are conserved in the SALT--Euler equations. Denoting the spatial $L^2$ inner product by angled brackets $\langle \cdot, \cdot\rangle$ and the using integration by parts identity for the Poisson bracket $\langle\{a, b\}, c\rangle = \langle\{b, c\}, a\rangle$, Casimir conservation is also easily verified via direct calculation,\begin{equation}
    \begin{split}
        \td C_n(\omega) = \langle\td\omega,\omega^{n-1}\rangle = \langle \{\psi\dt + \zeta_i\circ\dw, \omega\},\omega^{n-1}\rangle = \langle \underbrace{\{\omega, \omega^{n-1} \}}_{=0}, \psi\dt + \zeta_i\circ\dw \rangle = 0.
    \end{split}
\end{equation}
However, the energy is not conserved in the SALT approach due to the addition of noise Hamiltonians in the variational principle.

\paragraph{SFLT} The SFLT framework provides an alternative geometric stochastic forcing. 
Derived from a Lagrange--D'Alembert principle, this approach allows for the inclusion of arbitrary stochastic forcing within the governing equations for both momentum and advected quantities.
Specific energy-preserving stochastic forcings can thus be designed when the Poisson bracket governing the underlying dynamics is known. The corresponding SFLT--Euler equations take the following form: \begin{equation}
\td\omega + \{\psi, \omega\,\dt + \sum_i\theta_i\circ\dw\} = 0. 
\end{equation}

Here, $\theta_i=\theta_i(x)$ are fixed, space-dependent scalar fields and $W_t^i$ are independent Wiener processes. In the Lie--Poisson form, the equations are given by \begin{equation}
    \td \omega + \ad_{\frac{\delta H}{\delta \omega}}^*\left(\omega\,\dt + \theta_i\circ\dw\right)=0.
\end{equation}
The additional terms entering in the $\ad_\frac{\delta H}{\delta \omega}^*$ operator indicate that the Lie--Poisson structure is not preserved.
However, conservation of energy is readily verified using the properties of the Poisson bracket,
\begin{equation}
    \td H(\omega) = -\frac{1}{2}\langle \td\omega,\psi\rangle = -\frac{1}{2}\langle\{\psi, \omega\,\dt + \theta_i\circ\dw\}, \psi\rangle = -\frac{1}{2}\langle\underbrace{\{\psi, \psi\}}_{=0}, \omega\,\dt + \theta_i\circ\dw\rangle = 0.
\end{equation}
The SFLT and SALT frameworks differ since, in the former, stochasticity is introduced only after taking variations of the Hamiltonian. In the case of energy-preserving forcing, the Kelvin circulation theorem is modified and Casimir functions are generally no longer conserved.

\paragraph{LA SALT}
We include the LA SALT framework here to demonstrate its utility as a regularized alternative to SALT, without providing an exhaustive derivation (we refer to \cite{drivas2019lagrangian, alonso2020modelling} for more details).
In LA SALT, the advecting drift velocity $\gradperp\psi$ is replaced by its expectation \cite{drivas2019lagrangian}, and so distinguishes between `climate' and `weather' \cite{alonso2020modelling}. Namely, the `climate' is represented by the deterministic, expected evolution of the fluid while the `weather' is governed by linear stochastic equations that describe how fluctuations are transported by and interact with the mean flow. Furthermore, replacing the advecting velocity by its expectation introduces a non-locality in probability space that regularizes the dynamics \cite{drivas2019lagrangian}.

The LA SALT equations for two-dimensional Euler flow are given by \begin{equation}
    \td\omega + \{\E[\psi]\,\dt + \sum_i\zeta_i\circ\dw, \omega\} = 0, \label{eq:LA SALT}
\end{equation}
where $\zeta_i$ are defined as in the SALT framework. The energy is no longer conserved due to the change in the advecting velocity, however, the Lie--Poisson structure is the same as in the two-dimensional Euler equations. Equivalently, the velocity of material masses along a moving loop in the Kelvin circulation is altered, while the momentum per unit mass remains unchanged. Consequently, the LA SALT-Euler formulation possesses the same Casimir functionals as the deterministic two-dimensional Euler system and the SALT-Euler system.

The LA SALT expectation equation can be computed after passing to the \ito form of \eqref{eq:LA SALT}. We refer to \cite{alonso2020modelling} for more details. The conversion of the (LA) SALT equations from the Stratonovich form to the \ito form introduces a nested transport term \cite{holm2015variational}, after which taking the expectation yields \begin{equation}
    \frac{\partial}{\partial t}\E[\omega]+\{\E[\psi], \E[\omega]\} = \sum_i\frac{1}{2}\bxi_i\cdot\grad(\bxi_i\cdot\grad \E[\omega]), \label{eq:LA SALT_expectation}
\end{equation}
where $\bxi_i=\gradperp\zeta_i$.
The notation of the term on the right-hand side emphasizes its interpretation as a diffusive term. An alternative notation is as a nested Poisson bracket, $\frac{1}{2}\{\zeta_i, \{\zeta_i, \E[\omega]\}\}$, which highlights its relation to vorticity transport.

A distinct property of LA SALT is that the evolution of the expectation and the covariance tensor is closed\footnote{The evolution of the covariance of the variables in LA SALT does not generally form a closed system, see \cite{alonso2020modelling}.} \cite{drivas2019lagrangian, holm2023deterministic}. This raises the question whether the expectation and covariance equations can be solved as a computationally cheap proxy of a full stochastic ensemble. We do not endeavor to provide a conclusive answer to this question in the present paper, as this warrants separate analysis and computational studies. Instead, in the numerical investigations that follow, we evaluate LA SALT from two distinct viewpoints. First, we simulate a finite-size ensemble using \eqref{eq:LA SALT} where the expectation is replaced by the ensemble mean. Such a finite-size setting is common in practical uncertainty quantification and data assimilation. Second, we solve the expectation equation \eqref{eq:LA SALT_expectation} directly to assess the diffusion introduced by the stochastic transport in the evolution of the mean. We restrict our current investigation to its performance in isolated vortex interactions to establish its regularizing properties, reserving the study of its sensitivity in fully developed turbulent flow for future work.

\paragraph{EA SFLT}
The Eulerian-averaged SFLT (EA SFLT) framework provides a mean field approximation to SFLT via probabilistic averaging. We include this framework as a counterpart to LA SALT and demonstrate its regularizing features numerically. The resulting averaging of the Lie--Poisson bracket regularizes the dynamics via the induced non-locality in probability space,
\begin{equation}
    \td\omega + \left\{\psi, \E[\omega]\,\dt + \sum_i\theta_i\circ\dw\right\} = 0.
\end{equation}
Here, $\theta_i$ are the same as used in SFLT.
The EA SFLT equations retain the energy-preserving property of SFLT. Similar to LA SALT, an appealing feature of the averaged framework is that the expectation and fluctuations are governed by closed dynamics, thus allowing direct investigation of flow statistics.
The corresponding equation for the mean vorticity is given by
\begin{equation}
    \frac{\partial}{\partial t}\E[\omega] + \{\E[\psi], \E[\omega]\} = \frac{1}{2}\sum_{i}\left\{ \theta_i, \Delta^{-1}\left\{\theta_i, \E[\psi] \right\} \right\}.
    \label{eq:EA_SFLT_expectation}
\end{equation}
The EA SFLT approach will simulated in two ways.
First, as a finite-size ensemble, replacing the expectation in Eq. \eqref{eq:EA_SFLT_expectation} by the ensemble mean. Second, by solving the expectation equation \eqref{eq:EA_SFLT_expectation} directly.
As with LA SALT, an assessment of the sensitivity of EA SFLT in turbulent regimes is deferred for future work.

The (LA) SALT and (EA) SFLT approaches differ fundamentally due to their conservation properties. Namely, the former enters as a perturbation to the stream function (and thus to the advecting velocity) while the latter enters as a perturbation to the vorticity. Both methods are characterized by a specific multiplicative noise, acting either on the vorticity gradient in (LA) SALT or on the stream function gradient in (EA) SFLT.
The vorticity itself is the Laplacian of the stream function and therefore contains smaller-scale features and sharper gradients. Therefore, we expect (LA) SALT to exert a comparatively larger influence than (EA) SFLT when using an identical noise basis. We highlight this feature via two different perspectives below.

\subsection{Two different perspectives for comparing SALT and SFLT} \label{subsec:perspectives}
To compare the impact of SALT and SFLT, we analyze their behavior across both frequency and physical domains. We first establish the difference in sensitivity across wave numbers. This is followed by a comparison of the interaction with an idealized vortex, which demonstrates how the sensitivity manifests when interacting with localized spatial structures. 

\paragraph{Comparison in spectral space}
A first comparison between the approaches is made by assessing the interactions of Fourier modes introduced in the forcing. To that end, we denote by $F_\bk$ the Fourier mode corresponding to the wave vector $\bk=(k, l)^T$, \begin{equation}
    F_\bk = \exp(2\pi i \bk\cdot\bx),
\end{equation}
and analogously define the wave vector $\bm=(m, n)^T$ and the corresponding Fourier mode $F_\bm$.

It is evident that $\Nabla F_\bk = 2\pi i F_\bk \bk$ and $\Delta F_\bk = -4\pi^2 \|\bk\|^2 F_\bk$, where $\|\bk\|^2 =  k^2+l^2$. 
That is, the Fourier modes are eigenfunctions of the Laplacian on the torus where the eigenvalues depend on the wave number.
We may decompose the $\psi$ and $\omega$ into Fourier modes using the $L^2$ inner product $\langle\cdot,\cdot\rangle$, and denote $\psi_\bk:=\langle\psi,F_\bk\rangle F_\bk$ and $\omega_\bk:=\langle\omega, F_\bk\rangle F_\bk$.
Furthermore, via the relation $\Delta\psi=\omega$, we have that $\omega_\bk = -4\pi^2\|\bk\|^2\psi_\bk$.
With this in mind, we obtain the SALT term \begin{equation}
    \begin{split}
        \{F_\bm, \omega_\bk\} 
        & = \gradperp F_\bm\cdot\nabla\omega_\bk  \\
        & = -4\pi^2 F_\bm F_\bk\langle\omega, F_\bk\rangle\bm^\perp\cdot\bk \\
        & = 16\pi^4 \|\bk\|^2 F_\bm F_\bk\langle\psi, F_\bk\rangle \bm^\perp\cdot\bk
    \end{split}
    \label{eq:SALT_spectral}
\end{equation}
and the SFLT term
\begin{equation}
    \begin{split}
        \{\psi_\bk, F_\bm\} 
        & = \gradperp\psi_\bk\cdot\nabla F_\bm \\
        & = -4\pi^2F_\bm F_\bk\langle\psi, F_\bk\rangle \bk^\perp\cdot\bm \\
        &= 4\pi^2F_\bm F_\bk\langle\psi, F_\bk\rangle \bm^\perp\cdot\bk.
    \end{split}
    \label{eq:SFLT_spectral}
\end{equation}

\begin{remark}\label{Spectral-diff}
A comparison of Equations \eqref{eq:SALT_spectral} and \eqref{eq:SFLT_spectral} reveals that introducing a Fourier mode as a SALT term induces a larger effect than as a SFLT term.
The interaction with the vorticity in \eqref{eq:SALT_spectral} yields an amplified interaction with high-frequency flow components, evident from the factor $\|\bk\|^2$. Conversely, the interaction of the Fourier mode with low-frequency flow components is similar between the two approaches. 

Compared to SALT, the stream function averaging in LA SALT reduces high-frequency components within the advecting velocity. Thus, while the nonlinear interaction between the stream function and the vorticity does not change when passing from SALT to LA SALT, growth of high-frequency vorticity components is reduced. Consequently, a reduced interaction between stochastic terms and vorticity should produce a more condensed spread in ensemble simulations.
However, the magnitude of this reduction is challenging to estimate because it depends on the flow conditions and the ensemble prediction itself.
\end{remark}

\begin{remark}\label{rem:expectation_diff}
A similar distinction arises when comparing the averaged frameworks LA SALT and EA SFLT. In the mean equation for EA SFLT \eqref{eq:EA_SFLT_expectation}, the regularization term contains the expectation of the stream function, $\E[\psi]$, and the inverse Laplacian $\Delta^{-1}$. In contrast, the regularization in the LA SALT mean equation \eqref{eq:LA SALT_expectation} is governed by the expected vorticity $\E[\omega]$. Since the vorticity and stream function are related via $\Delta\psi=-\omega$, the former inherently possesses sharper gradients and more prominent high-frequency components. Following the reasoning of Remark \ref{Spectral-diff}, the diffusion induced by LA SALT is therefore expected to be more significant than that of EA SFLT. 
\end{remark}

\paragraph{Interaction with idealized vortex} 
As a second comparison, we study the interaction of the stochastic approaches with an idealized vortex that is axisymmetric around its center. This comparison permits an interpretation of sensitivity to high-frequency flow components in terms of coherent spatial structures. 

Letting the vortex center be located at the origin, a straightforward formulation of the vortex in polar coordinates can be found. 
We consider the two-dimensional plane and denote Cartesian coordinates by $x,y$ and polar coordinates by $r,\phi$. The transformation from Cartesian to polar coordinates is given by $r=\sqrt{x^2+y^2}$ for the distance to the origin and $\varphi=\atan(y, x)$ for the angle counterclockwise from the $x$-axis. Equivalently, one may use the relations $x=r\cos\varphi$ and $y=r\sin\varphi$.
The Laplacian in polar coordinates reads \begin{equation}
    \Delta = \frac{\partial^2}{\partial r^2} + \frac{1}{r}\frac{\partial}{\partial r} + \frac{1}{r^2}\frac{\partial^2}{\partial \theta^2}.
\end{equation}
Given a radial stream function, i.e., $\psi(r,\varphi)=\psi(r)$, it is easily verified from the definition of the Laplacian that $\omega=\Delta\psi$ is also radial.
Taking an arbitrary smooth function $f(r,\varphi)$, the Poisson bracket with a radial vorticity $\omega$ and stream function $\psi$ reduces to \begin{align}
    \{\psi, f\} &= \frac{1}{r}\frac{\partial f}{\partial \varphi}\frac{\partial\psi}{\partial r}, \label{eq:vortex_bracket0}\\
    \{f, \omega\} &= -\frac{1}{r}\frac{\partial f}{\partial \varphi}\frac{\partial\omega}{\partial r}=\{\psi, f\}\left(\frac{\partial\omega}{\partial r}\Big/\frac{\partial \psi}{\partial r}\right).
    \label{eq:vortex_bracket}
\end{align}
In the above derivation, we assume that the partial derivative of $\psi$ w.r.t. $r$ does not vanish.
Equation \eqref{eq:vortex_bracket} shows that the difference between SALT and SFLT on an ideal vortex is determined entirely by the ratio between the vorticity and stream function gradients. The choice of function $f$ does not alter this ratio, however, it affects the absolute difference between the two approaches.

The differences between \eqref{eq:vortex_bracket0} and \eqref{eq:vortex_bracket} can be made more explicit by assuming a specific stream function. 
As an example, we consider a coherent vortex structure described by radial Gaussian stream function, $\psi(r) = e^{-cr^2}$.
Here, the value of $c$ determines the size of the vortex and its decay rate in the radial direction, and is analogous to the wave number used for comparing in spectral space.
Using the definition of the Laplacian, straightforward computations yield \begin{align}
    \frac{\partial \psi}{\partial r} = -2cr e^{-cr^2}, &\quad
    \frac{\partial^2 \psi}{\partial r^2} = 2c e^{-cr^2}(2cr^2-1), \\
    \omega = 4c e^{-cr^2}(cr^2 - 1), &\quad
    \frac{\partial \omega}{\partial r} = -8c^2re^{-cr^2}(cr^2-2).
\end{align}

We assume now that the function $f$ is defined as the product of a radial function and sinusoidal function of the angle, $f(r,\varphi)=g(r)\sin(k\varphi)$. The function $g$ will not be specified while $k$ is an integer representing the wave number in the angular direction.
Substituting this into the Poisson bracket, we obtain 
\begin{equation}
    \begin{split}
        \{\psi, f\}(r, \varphi) &= -\frac{1}{r}g(r)k\cos(k\varphi)2cr e^{-cr^2} \\
        &= g(r)k\cos(k\varphi)2c e^{-cr^2}
    \end{split} \label{eq:ideal_vortex_bracket1}
\end{equation}
and
\begin{equation}
    \begin{split}
        \{f, \omega\}(r, \varphi) &= -\frac{1}{r}g(r)k\cos(k\varphi)\cdot(-8c^2re^{-cr^2}(cr^2-2)) \\
        &=g(r)k\cos(k\varphi)8c^2e^{-cr^2}(cr^2-2) \\
        &= -\{\psi, f\}\,4c(cr^2-2).
    \end{split}\label{eq:ideal_vortex_bracket2}
\end{equation}
It is readily verified that the ratio between \eqref{eq:ideal_vortex_bracket1} and \eqref{eq:ideal_vortex_bracket2} equals the ratio of the gradients of the vorticity and the stream function.

For this particular example, the ratio characterizing the difference between SALT and SFLT is given by $4c(cr^2-2)$ and grows quadratically with $r$. However, the vorticity and the stream function are exponentially decaying functions of $r$ and therefore the absolute difference between the two approaches vanishes in the limit of large $r$. The quadratic dependency on $c$ suggests that as vortices become smaller (represented by larger $c$), SALT grows significantly stronger than SFLT, resulting in an amplified effect on fine-scale features. The linear dependency on $k$ suggests sensitivity of the flow to high-frequency forcing components in both methods.

The comparisons of spectral components and the interaction with an idealized vortex suggest that SALT will exhibit higher sensitivity to fine-scale flow features than SFLT, a hypothesis that will be assessed numerically in the following section.

\section{Numerical experiments}\label{sec:numerical_experiments}
In this section, we study the effects of the stochastic forcings on the numerical solution in three test cases: the traveling dipole, vortex merger, and forced-damped turbulence. These scenarios are selected to systematically evaluate the physical behavior induced by the stochastic forcings and to identify the conditions under which each method is most effectively applied.

The traveling dipole and vortex merger test cases serve as controlled experiments that allow us to isolate the interaction of noise with localized vortex structures, providing a basis for comparing how SALT and SFLT differ in their treatment of advection and forcing.
By examining these simple interactions, we identify how the trade-off between preserving Casimirs or energy impacts the representation of uncertainty.
Subsequently, we apply these observations to a forced-damped turbulent flow, which serves as a demonstration of how the behaviors identified in the simpler test cases carry over to complex multi-scale flow.

We first introduce the computational method and the chosen stochastic terms before proceeding to the comparison of the numerical results.

\subsection{Numerical method}
Below, we mention the key aspects of the adopted numerical method. Appendix \ref{app:sec_numerical_method} provides a detailed description of the numerical method, including relevant references.

To discretize the system (\ref{eq:vorticity_advection}-\ref{eq:strm_vort_relation}) spatially, we employ a finite element approach for general vorticity dynamics \cite{bernsen2006dis} using a mixed continuous and discontinous Galerkin scheme.
Namely, the elliptic equation for the stream function \eqref{eq:strm_vort_relation} is discretized using a continuous Galerkin (CG) scheme. A discontinuous Galerkin (DG) scheme is used to discretize the transport equation \eqref{eq:vorticity_advection}, where upwinding is used to handle discontinuities across neighboring elements. The diffusive term arising in the LA SALT expectation equation \eqref{eq:LA SALT_expectation} is treated with a symmetric interior penalty Galerkin (SIPG) scheme. 
The nested Poisson bracket appearing in the EA SFLT expectation equation \eqref{eq:EA_SFLT_expectation} is solved numerically by introducing an auxiliary variable to compute the inverse Laplacian of the inner bracket.

In the absence of diffusion, the resulting scheme combined with a third-order explicit Runge--Kutta method preserves energy and is $L^2$-stable in the energy norm. We use an explicit integrator for its efficiency and straightforward adaptability to a stochastic settings. However, enstrophy is not preserved numerically.

A structured triangular mesh is used throughout all simulations, subdividing the domain in $256^2$ elements. The time step size is set to $\Delta t=5\times 10^{-3}$ time units and the simulations are run until $100$ time units have passed. Each stochastic ensemble consists of 10 members, which is found to provide a clear insight into the qualitative and quantitative differences between the stochastic approaches.

All computations are performed in the open-source Python finite-element package Firedrake \cite{FiredrakeUserManual}, which relies heavily on the PETSc toolkit for scientific computing \cite{petsc-user-ref}.

\subsection{Definition of stochastic terms}\label{subsec:stochastic_terms}
The spatial profiles that form the basis for the stochastic forcing are here chosen as two-dimensional Fourier modes. This offers the advantage of being well-interpretable and useful for discerning the specific effects of including different scales of motion in the stochastic forcing.
The expansion of the SALT-terms and the SFLT-terms in Fourier modes is given in Section \ref{subsec:perspectives} and indicates that the effect of SALT is generally larger than the effect of SFLT.
To ensure a fair comparison of the approaches, we choose the basis of the stochastic forcing as \begin{align}
    \zeta_\bk &= \sigma F_\bk \quad \text{(SALT)}, \label{eq:noise_terms_salt}\\
    \theta_\bk &= \sigma 4\pi^2 F_\bk \quad \text{(SFLT)} \label{eq:noise_terms_sflt}.
\end{align}
Here, $\sigma$ dictates the overall magnitude of the forcing and will be varied throughout the numerical experiments.
We select two groups of forcing based on their constituent wave numbers.
The first group comprises low-frequency modes satisfying $5\leq\|\bk\|$, while the second group consists of forcing in a band of higher frequencies $10\leq\|\bk\|\leq20$. In the subsequent figures, these frequency groups are respectively referred to as `lowfreq' and `highfreq'. 
In reduced-order stochastic modeling, the forcing terms can be informed by proper orthogonal decomposition (POD)\footnote{POD modes are often referred to as empirical orthogonal functions (EOFs) in the context of geophysical flows and as Karhunen-Lo\`eve basis functions in the field of stochastic processes. They are closely related to principal component analysis (PCA) and the singular value decomposition (SVD).}. In the present numerical tests, the low-frequency forcing emulates the energetic large-scale structures dominant in POD modes, whereas the high frequency forcing represents the smaller scales of motion that persist even in low-rank settings.

The stochastic forcing bases (\ref{eq:noise_terms_salt}-\ref{eq:noise_terms_sflt}) are suited for the torus.
The Fourier modes are multiplied by a mollified indicator function when applied to flows subject to the boundary condition \eqref{eq:bc}. The resulting spatial profiles are suited for bounded domains yet do not introduce numerically unresolvable gradients. The mollified indicator function adopted here is given by a product of logistic functions,
\begin{equation}
    \begin{split}
    \mathbb{1}_{\alpha}(x, y) &= 
    \left(1+\exp\left(\frac{6}{\alpha}(\alpha-x)\right)\right)^{-1}
    \left(1+\exp\left(\frac{6}{\alpha}(\alpha+x-1)\right)\right)^{-1} \\
    &\times
    \left(1+\exp\left(\frac{6}{\alpha}(\alpha-y)\right)\right)^{-1}
    \left(1+\exp\left(\frac{6}{\alpha}(\alpha+y-1)\right)\right)^{-1},
    \end{split}
    \label{eq:mollified_indicator}
\end{equation} where $\alpha$ can be considered the maximal resolvable wave number and is here chosen as $64$.

We remark that the choice of Fourier modes as a basis for the stochastic terms is essentially isotropic. As a result, the diffusion tensor appearing in the LA SALT expectation equation \eqref{eq:LA SALT_expectation_appendix} becomes spatially uniform in the domain, making the dissipation independent of the spatial coordinate.

\subsection{Traveling dipole}
The first test case concerns a traveling dipole, which provides a simple example of point-vortex dynamics for which an exact solution may be obtained \cite{zeitlin2018geophysical}. 
In the two-dimensional Euler equations, the dipole elongates in the direction of travel and separates when interacting head-on with a free-slip boundary  \cite{saffman1979approach}.
Numerically simulating a traveling dipole permits the study of the effect of noise on the interaction of two opposite-signed vortices, and serves as a controlled experiment of interactions of such vortices in turbulent flows.
In addition, it allows for the isolated study of the influence of stochasticity on coherent localized spatial structures and their advection.

\begin{figure}[htbp]  
    \centering
    \includegraphics[width=0.99\linewidth]{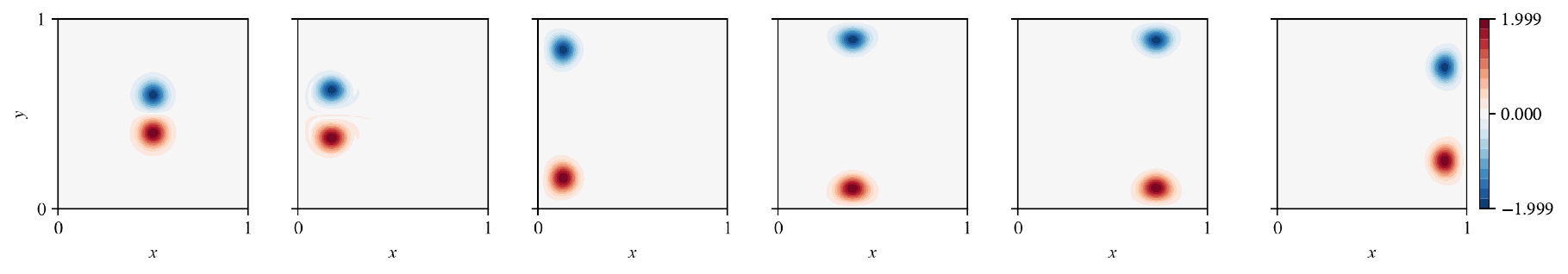}
    \caption{Deterministic evolution of the traveling dipole test case. From left to right, the vorticity snapshots are shown at times $t=0, 20, 40, 60, 80, 100.$}
    \label{fig:westward_evolution}
\end{figure}

In this numerical experiment the vorticity field is initialized as \begin{equation}
    \omega(x, y) = 2\left(e^{-\left((x-0.5)^2 + (y-0.35)^2\right)/0.005} - 
    e^{-\left((x-0.5)^2 + (y-0.65)^2\right)/0.005}\right)
\end{equation}
to generate a leftward moving dipole along the horizontal line $x=0.5$. The boundary condition \eqref{eq:bc} is used, causing the vortices to interact with the wall and subsequently split up.
The evolution of the deterministic solution is shown in Figure \ref{fig:westward_evolution}.

\begin{figure}[!ht]  
    \centering
    \includegraphics[width=0.99\linewidth]{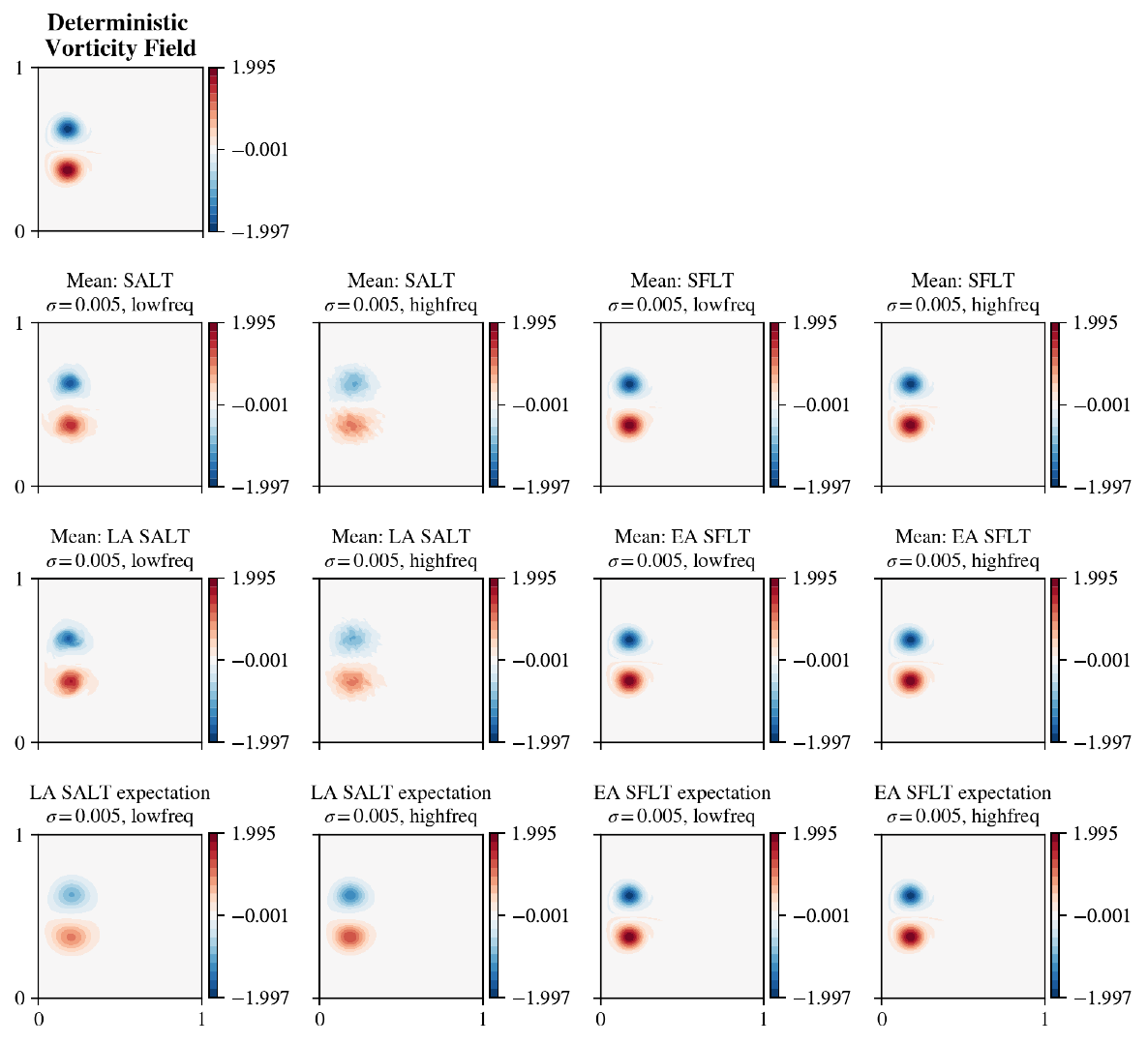}
    \caption{Visual comparison of the mean field for the traveling dipole test case. The deterministic vorticity field and the mean vorticity fields per ensemble are shown at time $t=20$. 
    The top row shows the SALT and SFLT ensemble means, the middle row shows the ensemble means of LA SALT and EA SFLT, and the bottom row shows the fields obtained by solving the corresponding deterministic expectation equations.
    Each ensemble consists of 10 independent realizations, the labels `lowfreq' and `highfreq' refer to the Fourier modes that comprise the stochastic forcing (defined in Section \ref{subsec:stochastic_terms}).}
    \label{fig:westward_mean}
\end{figure}

\begin{figure}[!ht]  
    \centering
    \includegraphics[width=0.99\linewidth]{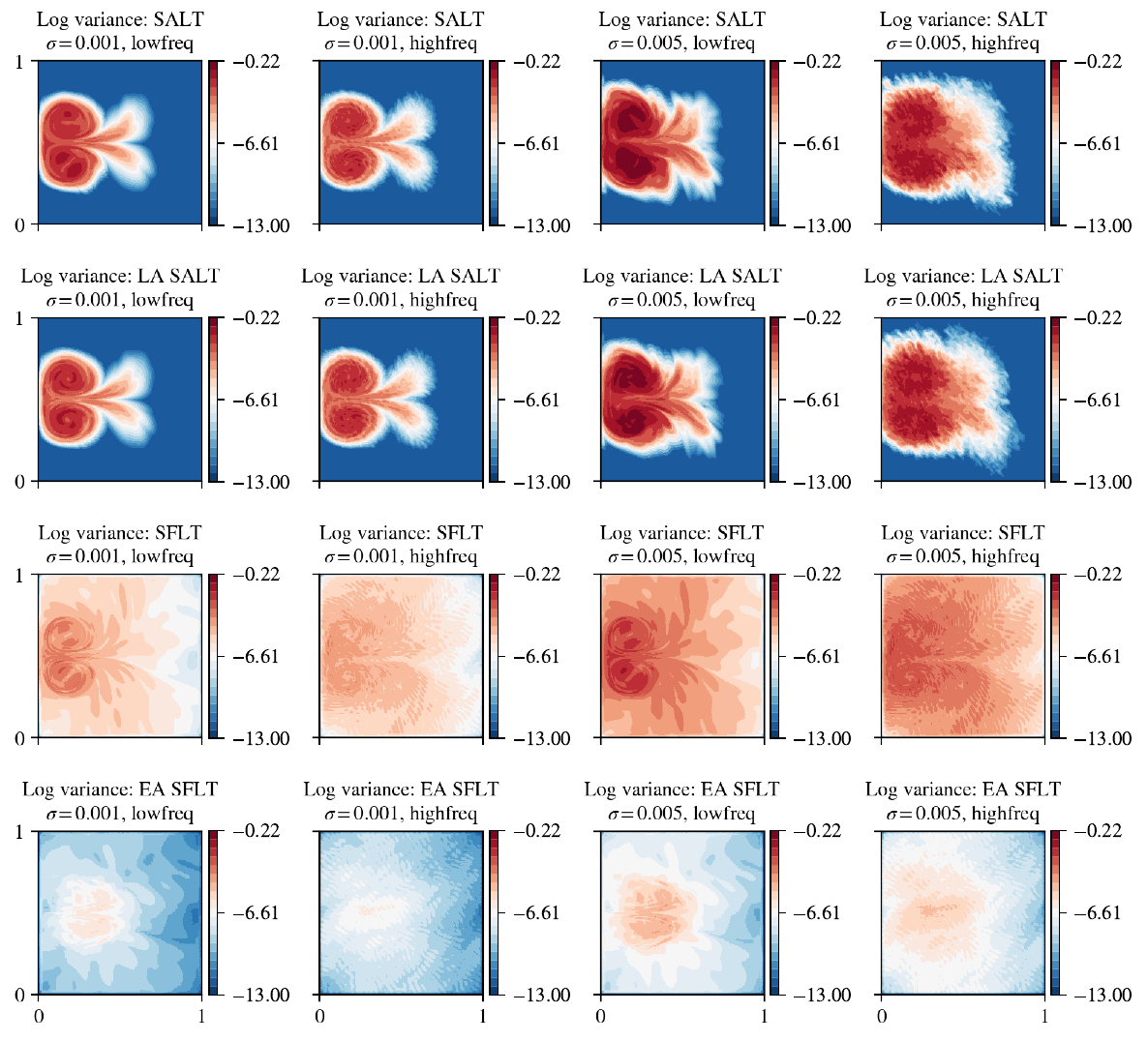}
    \caption{Visual comparison of the spatial variance for the traveling dipole test case at time $t=20$. The variance is shown per ensemble, each ensemble consists of 10 independent realizations, the labels `lowfreq' and `highfreq' refer to the Fourier modes that comprise the stochastic forcing (defined in Section \ref{subsec:stochastic_terms}). 
    Each row displays the results of a different stochastic framework, being (from top to bottom) SALT, LA SALT, SFLT, and EA SFLT.
    The common logarithm of the variance is used to highlight the qualitative differences between the stochastic methods.
    }
    \label{fig:westward_var}
\end{figure}

An intuitive overview of the qualitative differences between the two types of stochastic forcing is obtained by visualizing the mean and variance of the ensembles in space, at a fixed point in time. These fields are respectively shown in Figures \ref{fig:westward_mean} and \ref{fig:westward_var}, at time $t=20$. Several observations can be made from these figures.

Firstly, the mean vorticity fields suggest that the ensembles accurately capture the leftward movement of the dipole. However, a distinct spatial pattern is observed for the high-frequency (LA) SALT forcing, which is attributed to the enhanced effect of SALT on high-frequency vorticity components and small coherent structures (see Section \ref{subsec:perspectives}). The uncertainty modeled by the approaches, here quantified as the pointwise variance, reveals a clear distinction. By construction, (LA) SALT introduces uncertainty where nonzero vorticity gradients exist, resulting in a large variance concentrated closely around the dipole. 
As expected, this variance increases proportionally with the magnitude of the stochastic forcing.
Furthermore, both the LA SALT expectation solution and the LA SALT ensemble mean are found to be dissipative, and we attribute the difference between these solutions to the finite-size ensemble mean approximation.

In contrast, Figure \ref{fig:westward_var} shows that the uncertainty introduced by SFLT is present throughout a larger portion of the domain and is less sensitive to the precise location of the vortices. Its effect is generally smaller than that of SALT, yet is still seen to be concentrated around the dipole structure. This is a direct consequence of SFLT acting on the gradient of the stream function, which is smoother than the vorticity with smaller gradients, leading to uncertainty which is more broadly spread across the domain. 
The variance induced by EA SFLT is found to be substantially smaller than that of SFLT, caused by the averaging of the vorticity in the former (see also Remark \ref{rem:expectation_diff}).

\begin{figure}[!ht]  
    \centering
    \includegraphics[width=0.99\linewidth]{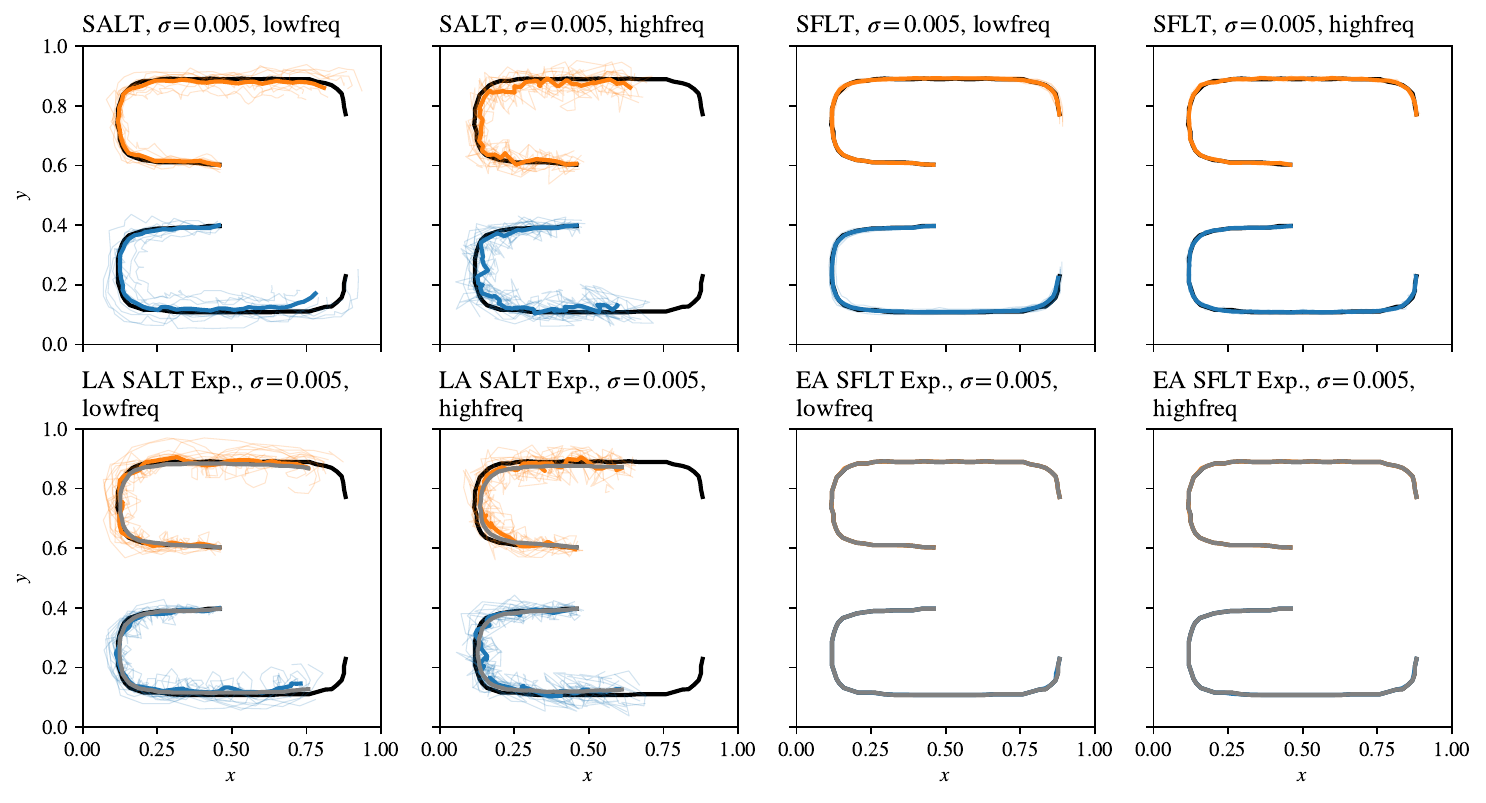}
    \caption{Vortex trajectories for the traveling dipole test case, showing the trajectories of the minimum value (blue line, bottom curve in each panel) and maximum value (orange line, top curve in each panel) of the vorticity from $t=0$ to $t=100$. The solid black lines depict the minimum and maximum value of the deterministic simulation and are shown for reference. 
    The top row shows the SALT and SFLT ensembles and their means, the bottom rows shows the LA SALT and EA SFLT ensembles and their means as well as the solution to the deterministic expectation equations.
    Each stochastic ensemble consists of 10 independent realizations, the labels `lowfreq' and `highfreq' refer to the Fourier modes that comprise the stochastic forcing (defined in Section \ref{subsec:stochastic_terms}). The thin solid colored lines show the trajectory of the individual ensemble members, the thick solid lines show the spatial mean of the trajectories. The trajectories corresponding to the LA SALT and EA SFLT expectation equations are depicted as thick gray curves only in the bottom row.
    }
    \label{fig:westward_trajectories}
\end{figure}

The vortex trajectories, obtained by tracing the maximum and minimum vorticity values, are depicted in Figure \ref{fig:westward_trajectories}. (LA) SALT induces a distinct change in these trajectories. Specifically, perturbations in the transport velocity cause the vortices travel a smaller net distance even as the mean trajectory continues to follow the deterministic trajectory. This effect is more pronounced under high-frequency forcing, reflecting the sensitivity of SALT to forcing at higher wave numbers. The deterministic LA SALT expectation, depicted in the two left panels in the bottom row in gray, accurately captures the mean trajectory of the corresponding ensemble. In contrast, trajectories from the (EA)SFLT simulations consistently mirror the deterministic reference trajectory. This qualitative behavior appears analogous to the dynamics of peakons in the stochastic Camassa--Holm equation, which exhibit spreading under SALT while remaining tightly clustered around the deterministic reference under SFLT \cite{holm2026comparative}.

\subsubsection{Traveling dipole with localized stochastic forcing}
To investigate the sensitivity of the stochastic parameterizations to the spatial distribution of noise, we repeat the traveling dipole case with stochastic forcing only active within the patch $(x,y)\in [0,0.4]\times[0, 0.5]$ by modifying the mollified indicator \eqref{eq:mollified_indicator} accordingly.
This provides a simplified setting for spatially varying stochastic forcing, for example obtained as eigenvectors of the velocity-velocity correlation tensor \cite{cotter2019numerically, ephrati2023noise}.
For brevity, we focus only on the SALT and SFLT frameworks.

\begin{figure}[!ht]
    \centering
    \includegraphics[width=0.99\linewidth]{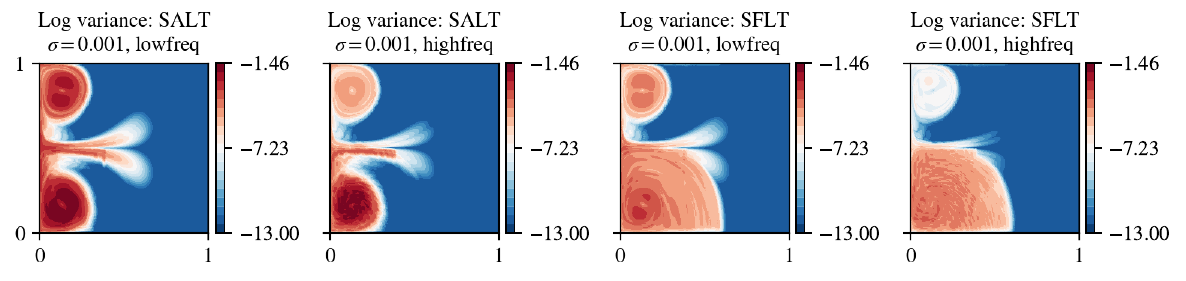}
    \caption{Visual comparison of the spatial variance for the traveling dipole test case at time $t=20$, using localized stochastic forcing near the bottom left corner of the domain. The variance is shown per ensemble, each ensemble consists of 10 independent realizations, the labels `lowfreq' and `highfreq' refer to the Fourier modes that comprise the stochastic forcing (defined in Section \ref{subsec:stochastic_terms}). The common logarithm of the variance is used to highlight the qualitative differences between the stochastic methods.}
    \label{fig:westward_partial_var}
\end{figure}

The spatial variance of the resulting ensembles at $t=20$ is visualized in Figure \ref{fig:westward_partial_var}.
As the dipole translates leftward, the lower vortex passes through the region of stochastic activity while the upper vortex remains outside the patch. Consequently, the noise directly imparts uncertainty on the former and indirectly on the later through the interaction between the vortices.
Consistent with previous results, SALT induces uncertainty concentrated around the vortex cores, while SFLT produces a more diffuse variance field that spreads out across the stochastic forcing patch. 
Furthermore, even as the vortices exit the stochastic patch and continue their trajectories, the ensemble members do not converge back to a single deterministic state. Instead, the imparted uncertainty persists, effectively acting as if it were a perturbation to the ensemble initial condition. This suggests that localized stochastic forcing can impact the predictability of coherent vortex structures even when they have moved to regions where no stochastic forcing is applied.

\clearpage
\subsection{Vortex merger}
\begin{figure}[htbp]  
    \centering
    \includegraphics[width=0.99\linewidth]{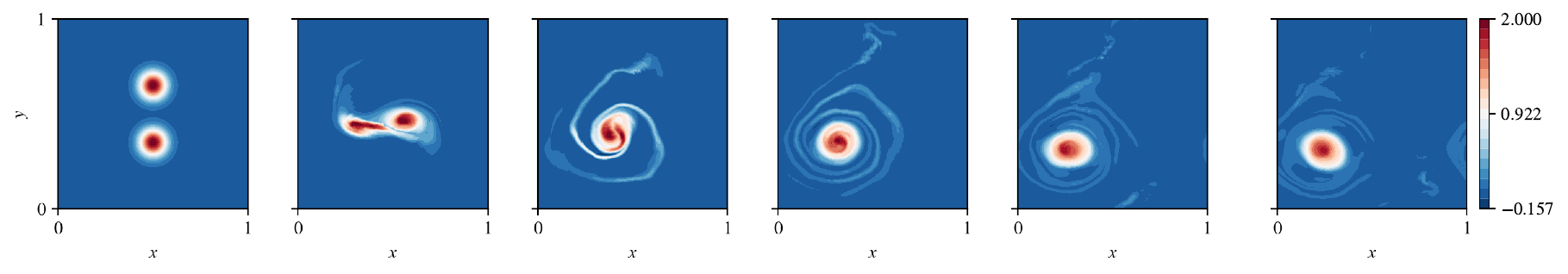}
    \caption{Deterministic evolution of the vortex merger test case. From left to right, the vorticity snapshots are shown at times $t=0, 20, 40, 60, 80, 100.$}
    \label{fig:merger_evolution}
\end{figure}

The second test case deals with the merger of like-signed vortices. While the traveling dipole illustrates how stochasticity affects the translation of coherent structures, the vortex merger test case allows us to examine the influence of noise on the self-organization and interaction of like-signed vortices. This self-organizing property is an important qualitative difference between two-dimensional and three-dimensional turbulence. Namely, like-signed vortices in two-dimensional turbulence tend to combine and form large-scale structures, driven by transfer of energy to low-wavenumber flow components \cite{chertkov2007dynamics}.
Here, we assess numerically the interaction of two vortices of the same sign and intensity to investigate the effects of stochastic forcing on vortex merger. We use an initial condition given by
\begin{equation}
    \omega(x, y) = 2\left(e^{-\left((x-0.5)^2 + (y-0.4)^2\right)/0.005} + 
    e^{-\left((x-0.5)^2 + (y-0.6)^2\right)/0.005}\right).
\end{equation}
The resulting deterministic solution over time is visualized in Figure \ref{fig:merger_evolution}.

\begin{figure}[htbp]  
    \centering
    \includegraphics[width=0.99\linewidth]{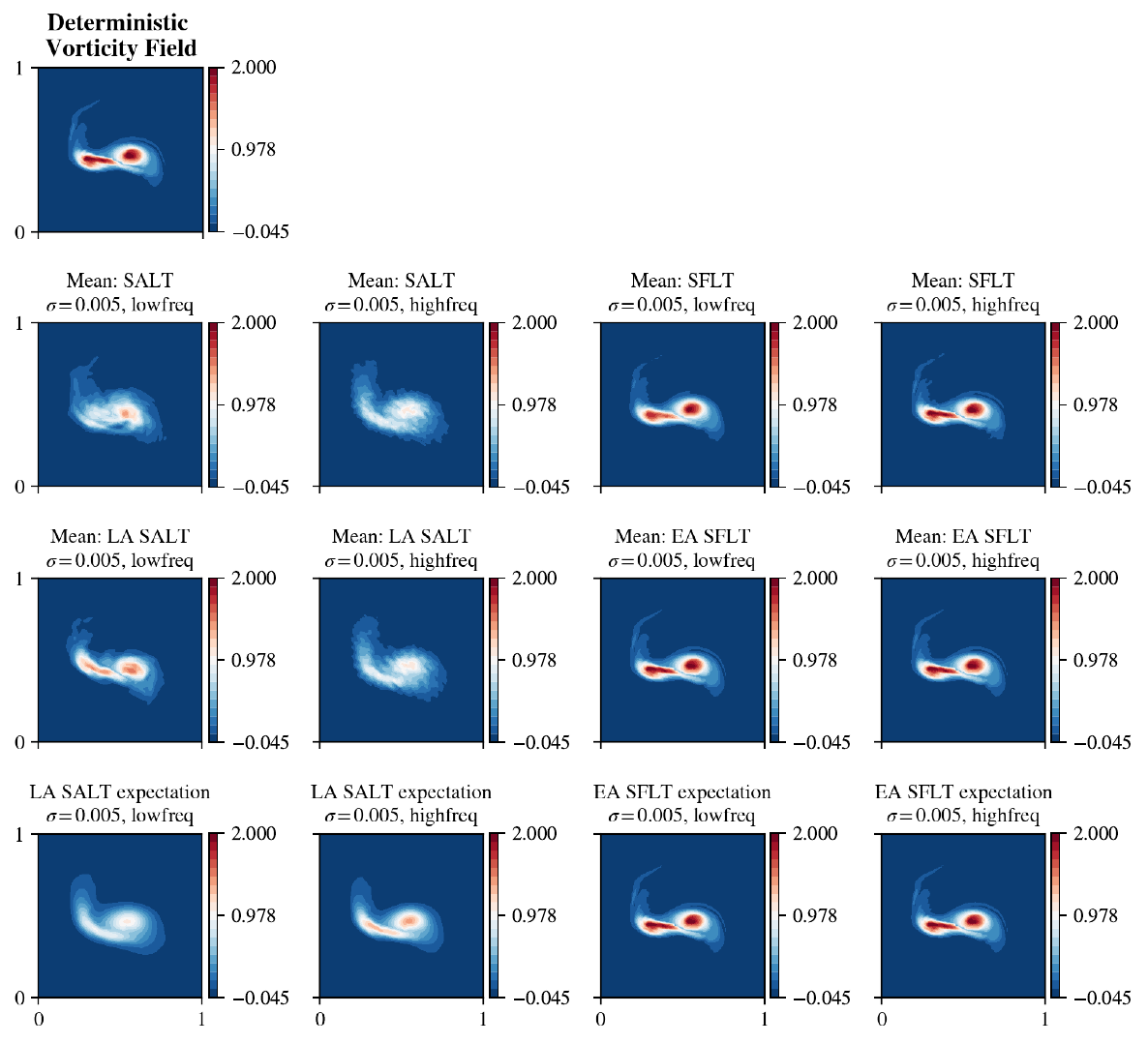}
    \caption{Visual comparison of the mean field for the vortex merger test case. The deterministic vorticity field and the mean fields per ensemble are shown at time $t=20$. 
    The top row shows the SALT and SFLT ensemble means, the middle row shows the ensemble means of LA SALT and EA SFLT, and the bottom row shows the fields obtained by solving the corresponding deterministic expectation equations.
    Each ensemble consists of 10 independent realizations, the labels `lowfreq' and `highfreq' refer to the Fourier modes that comprise the stochastic forcing (defined in Section \ref{subsec:stochastic_terms}).}
    \label{fig:merger_mean}
\end{figure}

\begin{figure}[!ht]  
    \centering
    \includegraphics[width=0.99\linewidth]{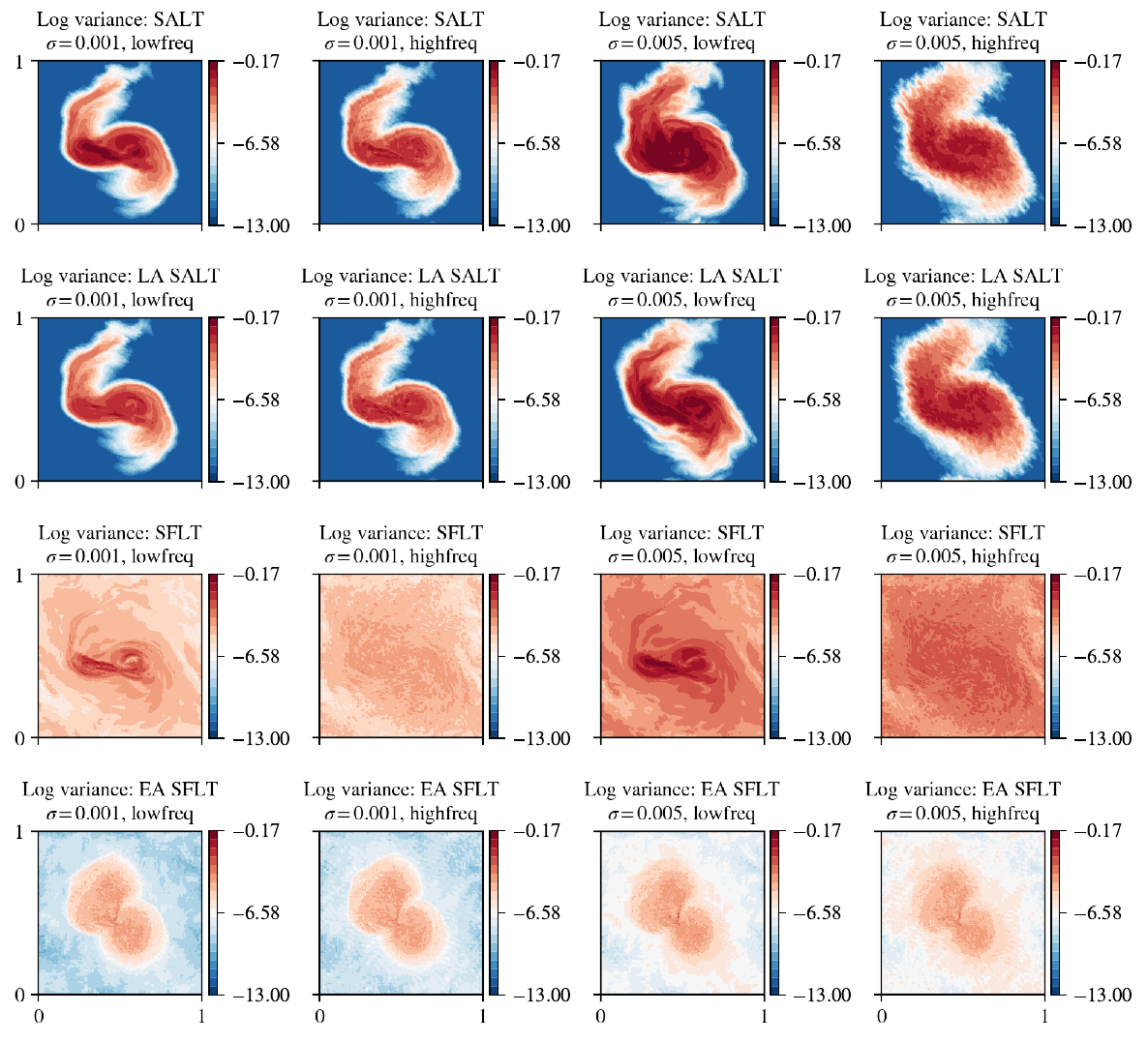}
    \caption{Visual comparison of the spatial variance for the vortex merger test case at time $t=20$. The variance is shown per ensemble, each ensemble consists of 10 independent realizations, the labels `lowfreq' and `highfreq' refer to the Fourier modes that comprise the stochastic forcing (defined in Section \ref{subsec:stochastic_terms}). 
    Each row displays the results of a different stochastic framework, being (from top to bottom) SALT, LA SALT, SFLT, and EA SFLT.
    The common logarithm of the variance is shown to highlight the qualitative differences between the stochastic methods. }
    \label{fig:merger_var}
\end{figure}

A visual comparison of the mean vorticity fields and ensemble variance fields is shown in Figures \ref{fig:merger_mean} and \ref{fig:merger_var}. Consistent with the traveling dipole case, the mean fields agree well with the deterministic solution. Both SALT and LA SALT yield a dispersed mean field, visible at both the low-frequency and high-frequency forcing, closely corresponding to the deterministic LA SALT expectation. 
The effect of (EA) SFLT on the vortex merger is less pronounced and no qualitative differences between the results can be discerned.
As before, the variance fields suggest that (LA) SALT models uncertainty close to the vortex structures while (EA) SFLT induces fluctuations in the vorticity that are spread across the domain. The variance of EA SFLT is found to be significantly smaller than that of SFLT, in line with the observations of the previous test case.

\begin{figure}[htbp]  
    \centering
    \includegraphics[width=0.99\linewidth]{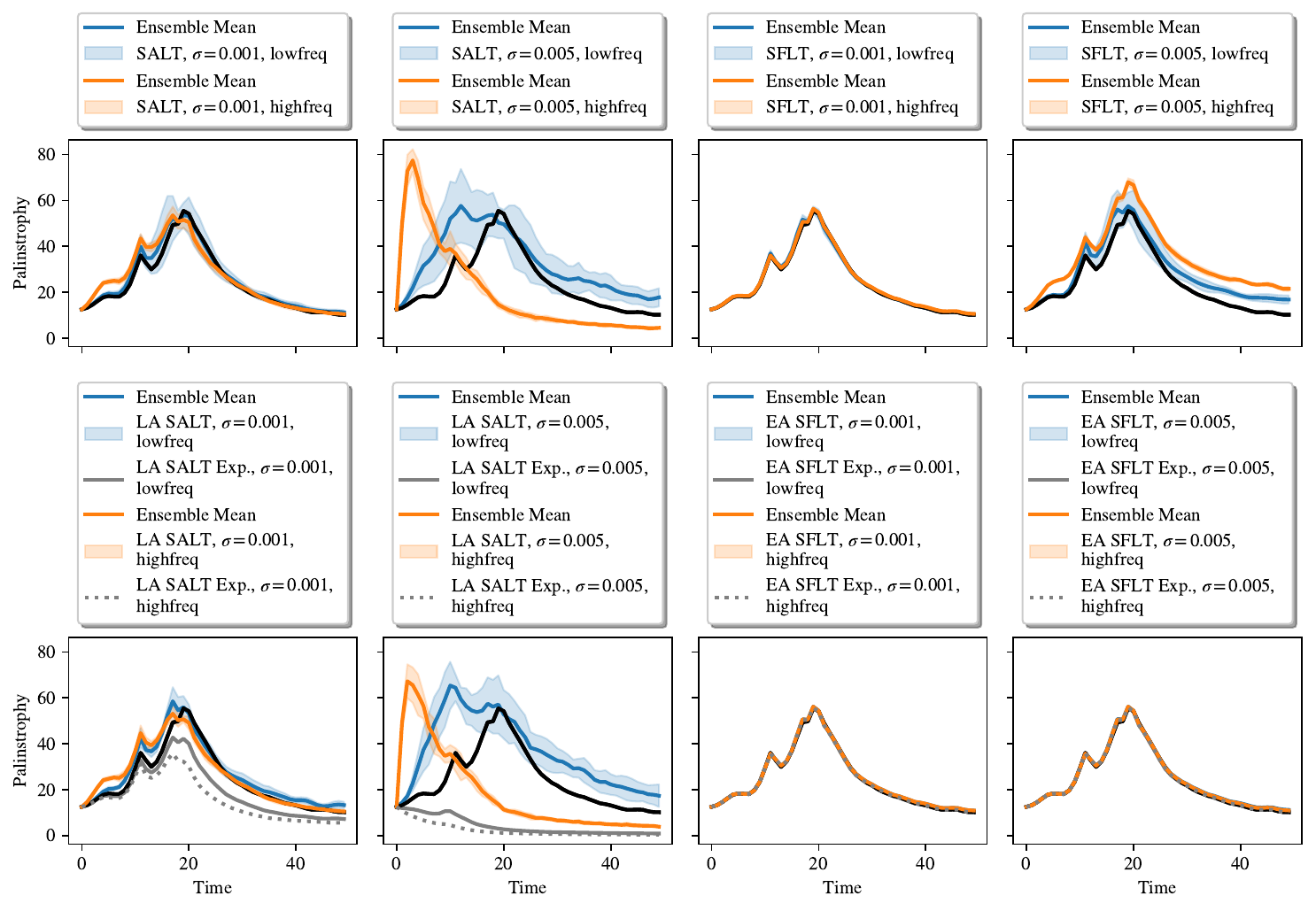}
    \caption{Comparison of the palinstrophy. The black lines show the values of the deterministic simulation. 
    The top row shows SALT and SFLT, the bottom row shows LA SALT and EA SFLT.
    For each ensemble, the solid line indicates the mean while the shaded areas denote one standard deviation around the mean. The values corresponding to the LA SALT and EA SFLT expectations are depicted in gray.}
    \label{fig:merger_palinstrophy}
\end{figure}

Figure \ref{fig:merger_palinstrophy} shows the palinstrophy over time, where each stochastic ensemble is compared to the reference deterministic simulation. The gradient of the vorticity in the definition of the palinstrophy (Equation \eqref{eq:palinstrophy} in Appendix \ref{app:qois}) ensures that the quantity is sensitive to the formation and presence of high-frequency components in the numerical solution.
As such, it provides quantitative insight into the appearance of vorticity filaments and small-scale structures. 
The deterministic solution (black line) shows palinstrophy increasing during the merger and subsequently decaying once the vortices have merged.

We observe that low-frequency SALT forcing with $\sigma=0.001$ is able to track the deterministic palinstrophy within one standard deviation of the ensemble mean. The high-frequency forcing at the same magnitude yields a similar palinstrophy development but with a reduced spread. For $\sigma=0.005$, the palinstrophy attains its peak value at an earlier time. This shift is driven by the high-amplitude and small-scale stochastic advection of the vorticity, which accelerates the break-up of vortices and the immediate formation of vorticity gradients. No qualitative differences are observed between the SALT and LA SALT ensembles. The latter shows slightly reduced spread, attributed to the regularization provided by advecting the vorticity with the ensemble mean drift velocity. The solution to the deterministic LA SALT expectation equation displays the effect of the dissipative term, particularly evident at when employing the high-frequency modes. The discrepancy between the LA SALT ensemble mean and the expectation stems from the fact that the palinstrophy of the expected vorticity field, representing a statistic of the mean, is not identical to the expected palinstrophy across the ensemble realizations, being the mean of a statistic.

The palinstrophy spread is narrower for SFLT than for SALT, as the latter interacts more strongly with high-frequency vorticity components. The effect of EA SFLT on the palinstrophy is nearly indistinguishable.Generally, SFLT at low frequencies is still found to accurately reproduce the deterministic palinstrophy. However, the final palinstrophy predicted by SFLT is higher than the deterministic value due to the introduced vorticity perturbations that increase the square of the vorticity gradient. 

\begin{figure}
    \centering
    \includegraphics[width=0.99\linewidth]{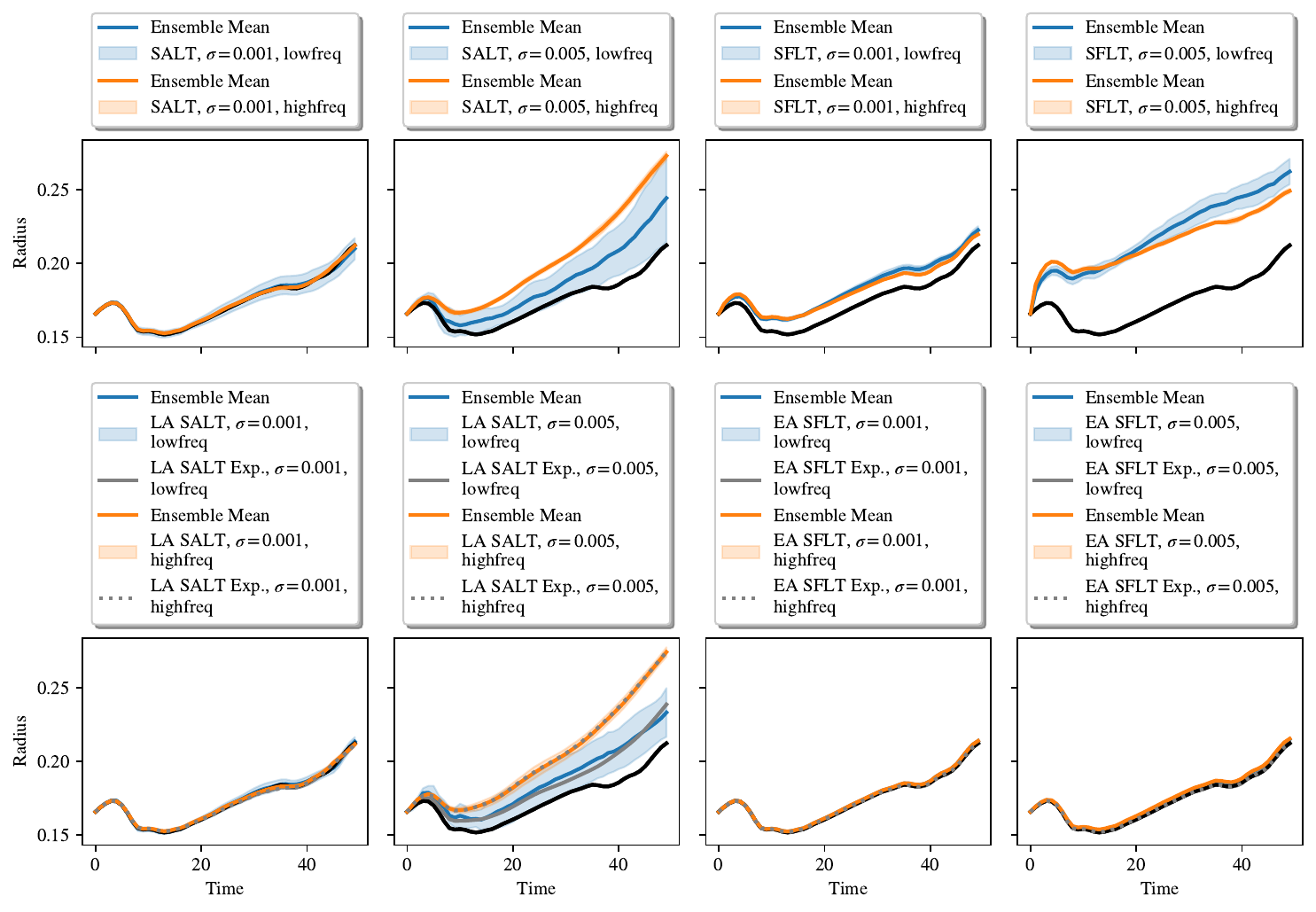}
    \caption{Comparison of the effective radius. The black lines show the values of the deterministic simulation. 
    The top row shows SALT and SFLT, the bottom row shows LA SALT and EA SFLT.
    For each ensemble, the solid line indicates the mean while the shaded areas denote one standard deviation around the mean. The values corresponding to the LA SALT and EA SFLT expectations are depicted in gray.}
    \label{fig:merger_radius}
\end{figure}

Figure \ref{fig:merger_radius} shows the effective radius relative to the positive center of mass (see \eqref{eq:effective_radius} in Appendix \ref{app:qois}). This quantity is the squared vorticity weighted by its distance from the center of mass and serves as measure of the spatial variance of the vorticity field. The deterministic evolution exhibits an initial decrease in the effective radius during the merger, followed by an increase after the two vortices have merged. At $\sigma=0.001$, SALT effectively follows the deterministic reference, with the low-frequency forcing providing an uncertainty estimate that accurately encompasses the reference. However, at $\sigma=0.005$, the SALT solutions deviate from the reference, suggesting that the vorticity realizations are less coherent than their deterministic counterpart. The same behaviour is observed for LA SALT, where the deterministic expectation provides a good estimate of the ensemble mean.
In the case of SFLT, the radius is generally larger than the reference, a result of vorticity perturbations distributed across the domain, whereas the effect of EA SFLT on the radius is negligible.

\clearpage
\subsection{Forced-damped two-dimensional turbulence}
As a final test case, we qualitatively compare SALT and SFLT in two-dimensional forced-damped turbulence.
This is realized by including additional terms to the governing equations, so that the flow is driven to a nontrivial statistically stationary state.
The vorticity evolution reads \begin{equation}
    \partial_t \omega = \{\psi, \omega\} + Q - r\omega,
\end{equation}
where $Q(x, y) = 0.1\sin(8 \pi x)$ and $r=0.01$, following \cite{cotter2019numerically, ephrati2023noise}. The equations are solved on a unit square with free-slip boundary conditions. 
The inclusion of stochasticity is not altered by the external forcing and damping terms. For this specific test case, we adopt a triangulation consisting of $512^2$ cells to increase the resolvable level of detail in the turbulent flow. The stochastic forcing is the same as described in the previous subsections, employing the mollified indicator \eqref{eq:mollified_indicator} to satisfy the boundary conditions. Only the noise magnitude $\sigma=0.001$ is used for brevity.

\begin{figure}[htbp]
    \centering
    \includegraphics[width=0.5\linewidth]{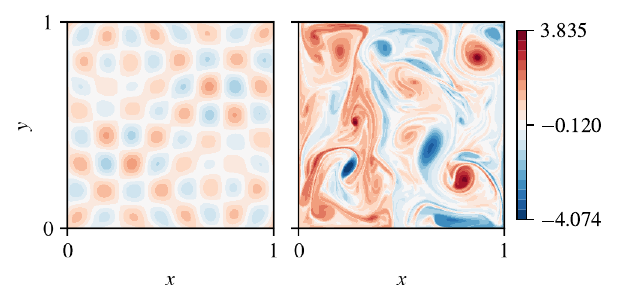}
    \caption{Left: Initial vorticity \eqref{eq:forced_damped_ic} for the forced-damped turbulence test case, before the simulation is spun up. Right: Vorticity fields after 200 time units, which defines the initial condition for the stochastic ensembles.}
    \label{fig:forced_damped_spinup}
\end{figure}

The flow is initialized from a prescribed vorticity field $\omega_0$, \begin{equation}
    \omega_0 = \sin(8\pi x)\sin(8\pi y) + 0.4\cos(6\pi x)\cos(6\pi y)+0.3\cos(10\pi x)\cos(4\pi y)+0.02\left(\sin(2\pi y)+\sin(2\pi x)\right), 
    \label{eq:forced_damped_ic}
\end{equation}
after which the simulation is spun up for 200 time units. The last stored vorticity field serves as an initial condition for the stochastic ensembles. The vorticity fields before and after the spin-up period are shown in Figure \ref{fig:forced_damped_spinup}.

While LA SALT and EA SFLT were evaluated in the previous test cases, its comparison is omitted in the turbulent regime. This choice is made to reserve the sensitivity of these methods to small-scale fluctuations for future work, focusing the current investigation on the differences between SALT and SFLT in the turbulent regime.

\begin{figure}[!ht]
    \centering
    \includegraphics[width=0.99\linewidth]{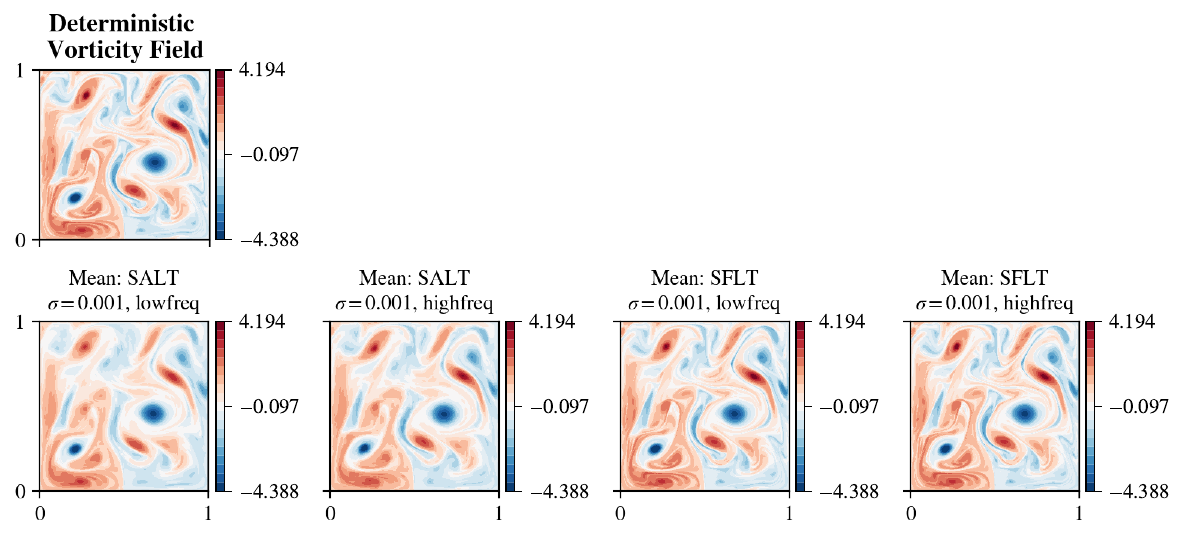}
    \caption{Visual comparison of the mean field for the forced-damped turbulence test case. The deterministic vorticity field and the mean vorticity fields per ensemble are shown 15 time units after the spin-up time. Each ensemble consists of 10 independent realizations, the labels `lowfreq' and `highfreq' refer to the Fourier modes that comprise the stochastic forcing (defined in Section \ref{subsec:stochastic_terms}).}
    \label{fig:box_mean}
\end{figure}

\begin{figure}[!ht]
    \centering
    \includegraphics[width=0.99\linewidth]{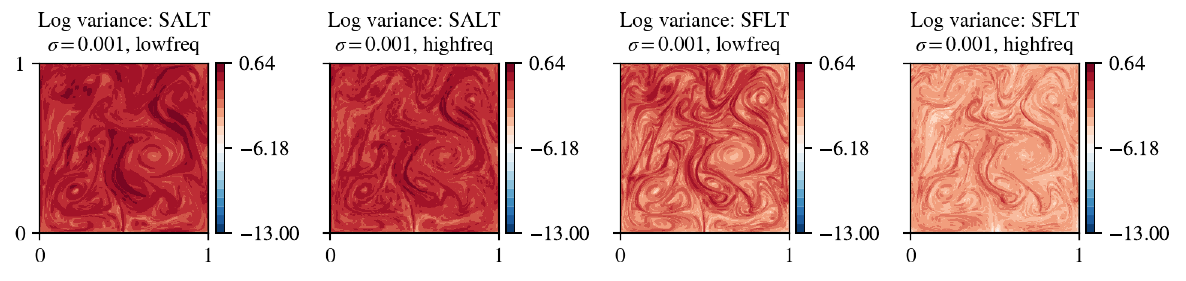}
    \caption{Visual comparison of the spatial variance for the forced-damped turbulence test case, shown at 15 time units after the spin-up time. The variance is shown per ensemble, each ensemble consists of 10 independent realizations, and the labels `lowfreq' and `highfreq' refer to the Fourier modes that comprise the stochastic forcing (defined in Section \ref{subsec:stochastic_terms}). The common logarithm of the variance is used to highlight the qualitative differences between the stochastic methods.}
    \label{fig:box_var}
\end{figure}

The qualitative differences between the forcing types can be seen from the ensemble means in Figure \ref{fig:box_mean} and the spatial variances in Figure \ref{fig:box_var}, compared at 15 time units after the spin-up.
The deterministic reference exhibits intricate patterns of both coherent vortices and elongated filaments.
For both SALT ensembles, the mean field appears significantly smoother, indicating that the transport noise induces variability in the positions of the vortex structures among the ensemble members.
In contrast, the SFLT mean fields remain sharp and closely resemble the deterministic reference. This suggests that, at the same magnitude, individual SLFT realizations remain more closely aligned to the deterministic reference than SALT.

The spatial distribution of uncertainty, shown in Figure \ref{fig:box_var} by the variance, is consistent with the previous test cases. Namely, in this turbulent regime SALT introduces variance that is localized near the edges of vorticity patches. 
Conversely, the SFLT variance is significantly lower in magnitude, evident from the lighter shading. In this turbulent flow, the variance induced by SFLT is found to also concentrate near vortex structures. The latter effect decreases as higher-frequency components are included in the stochastic forcing.

These results align with the differences between SALT and SFLT identified in the previously presented controlled settings. 
Namely, the locality of SALT and the increased modeled variance persist in fully developed turbulent flows. 
Consequently, it risks over-dispersing the mean field if the noise is not carefully calibrated. On the other hand, SFLT appears to provide a more moderate uncertainty estimate more equally spread across the domain.

\section{Conclusions}\label{sec:conclusions}
In this work, we have compared two geometric stochastic forcings within the context of vortex dynamics in the two-dimensional Euler equations. These are stochastic advection by Lie transport (SALT), which preserves the Lie--Poisson structure and Casimir functions, and stochastic forcing by Lie transport (SFLT), which preserves kinetic energy.
Through analysis of Fourier interactions and idealized vortices, validated by numerical tests, we have identified key distinctions in how these methods model uncertainty.

Our results demonstrate that the SALT framework introduces uncertainty predominantly in `dynamically active' regions where vorticity gradients are large, particularly visible from the spatial variances. This is a direct consequence of SALT acting as a perturbation to the velocity that advects the vorticity. 
This locality is both a strength and a challenge. Namely, it suggests that SALT is a good choice for uncertainty quantification in simulations where coherent flow structures are accurately predicted, while the palinstrophy and vortex trajectories show that it is significantly more sensitive to high-frequency flow components than SFLT. This emphasizes the need for accurate calibration of the stochastic forcing basis, without which SALT may induce non-physical effects such as the accelerated break-up of small vortices.

Conversely, the SFLT framework distributes uncertainty more broadly across the domain. By interacting with the stream function gradient, it produces stochastic fluctuations that are more evenly spread and less concentrated around specific spatial structures.
While spatial variances show that SFLT may be less precise in capturing uncertainty at sharp interface, it is a robust option when the precise location or existence of coherent spatial structures is not well-known.

The comparison presented in this paper may help refining calibration methodologies for sub-grid scale modeling and uncertainty quantification. Possible extensions of this work involve numerical studies of the two stochastic frameworks in increasingly complex systems, such as the (thermal) shallow water equations and ideal magnetohydrodynamics, serving as a necessary step in discerning the distinct physical behaviors induced by geometric noise in geophysical and plasma-physical flows.

A key question remains whether LA SALT, with closed equations for the expectation and fluctuations, can serve as a sufficiently accurate, low-cost surrogate for full stochastic ensembles.
Furthermore, examining these methods from the viewpoint of statistical physics, i.e., regarding invariant measures and behavior when combined with specific dissipation types, is essential for furthering the understanding of geometric stochasticity.

\section*{Acknowledgments}
The authors are grateful to Ruiao Hu and Oliver Street for insightful discussions during the preparations of this work.
DH was partially supported during the present work by Office of Naval
Research (ONR) grant award N00014-22-1-2082, Stochastic Parameterisation of Ocean Turbulence for Observational Networks (SPOT-ON) and by European Research Council (ERC)
Synergy grant Stochastic Transport in Upper Ocean Dynamics (STUOD) -- DLV-856408.
SE was partially supported during the present work by the Air Force Office of Scientific Research (AFOSR) grant award  FA8655-25-1-7465, Stochastic Plasma Physics Dynamics (SPPD) awarded to DH.

\appendix
\section{Numerical method}\label{app:sec_numerical_method}

The governing equations are discretized via a mixed finite element scheme. Sections \ref{app:subsec_strm_vort} and \ref{app:subsec_vort_advection} summarize the approach presented by \cite{bernsen2006dis, cotter2019numerically} for general vorticity dynamics. The treatment of the regularization terms in the LA SALT and EA SFLT expectation equations are detailed in Sections \ref{app:subsec_LA SALT} and \ref{app:subsec_EA_SFLT}, respectively, and time integration is discussed in Section \ref{app:subsec_time_integration}.

We let $\MT_h$ denote a triangulation of the domain $\Omega$ with maximum triangle diameter $h$. Let $K\in\MT_h$ be an element with boundary $\partial K$ and outward unit normal $\bn$. The spatial inner product over $K$ is denoted by $\langle\cdot, \cdot\rangle_K$ and a similar notation is used for inner products on the boundary. The space of continuous polynomials of degree at most $k$ on each element $K$ is denoted by $\MP^k(K)$.

\subsection{Elliptic stream function equation}\label{app:subsec_strm_vort}
The relation between the stream function and the vorticity \eqref{eq:strm_vort_relation} is discretized using a continuous Galerkin (CG) scheme. Let $W^1(\Omega)$ denote subspace of the Sobolev space $W^{1,2}(\Omega)$ with the same boundary conditions as the stream function. That is, no restrictions are imposed when the domain is periodic, and $\nu\in W^1$ satisfies $\|\nu\|_{\partial\Omega}=0$ when the domain is not periodic; the resulting discretization is the same for both settings.
Then, for each element $K$, we multiply equation \eqref{eq:strm_vort_relation} by
a test function $\phi\in\MW_h^k=\{\phi\in W^1(\Omega): \phi\in \mathcal{C}(\omega), \phi|_K \in\MP^k\}$ and integrate by parts to find \begin{equation}
    \langle\phi, \omega\rangle_K = -\langle\grad\phi, \grad\psi\rangle_K.
\end{equation}

\subsection{Vorticity advection equation}\label{app:subsec_vort_advection}

Throughout, we will use denote the velocity by $\bu$ instead of $\gradperp\psi$ for readability.

The vorticity advection equation \eqref{eq:vorticity_advection} is discretized in using a discontinuous Galerkin (DG) scheme. We denote by $p$ any test function in the space of discontinuous test functions $\MV_h^k = \{p\in K: \exists \phi_h\in\MP^k(K) \text{ s.t. } p|_K = \phi_h\}$.
Multiplying the vorticity equation by $p$ and integrating by parts yields \begin{equation}
    \langle p, \partial_t\omega\rangle_K = \langle\omega\bu, \nabla p\rangle_K - \langle\omega\bu\cdot\bn, p\rangle_{\partial K}.
    \label{eq:weak_form}
\end{equation}
The DG scheme ensures that $\omega$ and $p$ are discontinuous across elements. Therefore, a unique numerical flux must be defined to include the boundary integral appearing in the above equation. We adopt the following upwind scheme \cite{bernsen2006dis, cotter2019numerically}.

For points $\bx\in\partial K$, we define $p^{-}:=\lim_{\varepsilon\uparrow 0} p(\bx+\varepsilon\bn)$ and $p^{+}:= \lim_{\varepsilon\downarrow 0} p(\bx+\varepsilon\bn)$. The values $\omega^+$ and $\omega^-$ are defined analogously. Then, the numerical flux $\hat{f}$
\begin{equation}
    \hat{f}(\omega^+, \omega^-, \bu\cdot\bn) = \bu\cdot\bn\begin{cases}
        \omega^+ & \text{if $\bu\cdot\bn<0$} \\
        \omega^- & \text{if $\bu\cdot\bn\geq 0$}
    \end{cases}
\end{equation}
replaces $\omega\bu\cdot n$ in \eqref{eq:weak_form}. Simultaneously, $p$ in the boundary integral in \eqref{eq:weak_form} is replaced by $p^-$.

\subsection{LA SALT expectation equation}\label{app:subsec_LA SALT}

The expectation equation for LA SALT can be written as \begin{equation}
    \frac{\partial \E[\omega]}{\partial t} + \{\E[\psi], \E[\omega]\} = \grad\cdot(\BD\grad\E[\omega]). \label{eq:LA SALT_expectation_appendix}
\end{equation} 
The expectation $\E[\cdot]$ is omitted in the notation below, the variables $\psi, \omega$ are used instead. Equation \eqref{eq:LA SALT_expectation_appendix} consists of an advective term and a diffusive term, and is accompanied by an elliptic equation for the stream function. The vorticity advection is discretized following the method in Appendix \ref{app:subsec_vort_advection}, while the stream function is obtained from the vorticity followin the approach in Appendix \ref{app:subsec_strm_vort}. Details on the discretization of the diffusive term are given below. 

Recall that $\bxi_i=\gradperp\zeta_i$, where the latter are perturbations to the stream function. Then, the space-dependent diffusion tensor $\BD$ is defined as \begin{equation}
    \BD(\bx) = \frac{1}{2}\sum_i \bxi_i(\bx)\otimes\bxi_i(\bx).
\end{equation}
To ensure a stable and consistent discretization of the diffusion term $\grad\cdot(\BD\grad\omega)$ in conjunction with the DG method, a symmetric interior penalty Galerkin (SIPG) method (see, e.g., \cite{douglas2008interior, houston2002discontinuous}) is used. 
For an element $K$, let $\{\cdot\}$ and $\llbracket \cdot \rrbracket$ denote the average and jump operators on the boundary $\partial K$, defined respectively as
\begin{equation}
    \{ v \} = \frac{1}{2}(v^- + v^+) \quad \text{and} \quad \llbracket v \rrbracket = v^- - v^+.
\end{equation}
The discrete bilinear form associated with the diffusion operator is defined by summing the contributions of all elements.
The contribution of a single element $K$ to the bilinear form is given by:
\begin{equation}
    A_K(\omega, p) = \langle \grad p, \BD \grad \omega \rangle_K - \frac{1}{2} \langle \{\BD \grad \omega\} \cdot \bn, p \rangle_{\partial K} - \frac{1}{2} \langle (\BD \grad p) \cdot \bn, \llbracket \omega \rrbracket \rangle_{\partial K} + \langle \frac{\rho}{h} \{\BD\} \llbracket \omega \rrbracket, p \rangle_{\partial K},
\end{equation}
where $\rho > 0$ is a penalty parameter and $h$ is the local mesh size.  In this formulation, the first term represents the interior dissipation. The second and third terms ensure the consistency and symmetry of the operator, while the final term penalizes discontinuities in $\omega$ across element facets to ensure numerical stability.


\subsection{EA SFLT expectation equation} \label{app:subsec_EA_SFLT}
The expectation equation for EA SFLT is given by Eq. \eqref{eq:EA_SFLT_expectation}.
Unlike the LA SALT expectation equation, which features a standard (potentially anisotropic) diffusion term, the EA SFLT framework introduces a nonlocal regularization term involving an inverse Laplacian in the nested Poisson bracket.
The expectation equation reads \begin{equation}
    \frac{\partial}{\partial t}\E[\omega] + \{ \E[\psi], \E[\omega]\} = \frac{1}{2}\sum_i \left\{\theta_i, \Delta^{-1} \{\theta_i, \E[\psi]\} \right\},
    \label{eq:easflt_expectation_appendix}
\end{equation}
where $\theta_i$ are fixed spatial profiles that define the perturbations added to the vorticity in the SFLT framework.

To discretize the right-hand side of Eq. \eqref{eq:easflt_expectation_appendix}, we employ the mixed finite element method outlined in Sections \ref{app:subsec_strm_vort} and \ref{app:subsec_vort_advection}.
For each substage of the Runge--Kutta method (described in Section \ref{app:subsec_time_integration}), the regularization term is computed through the following sequence.
For each mode $\theta_i$, we compute the inner Poisson bracket with the mean stream function, $b_i:=\{\theta_i, \E[\psi]\}$. This bracket is evaluated as function in the DG space and subsequently interpolated to the CG space. We solve for an auxiliary variable $w_i$ satisfying $\Delta w_i = b_i$, following the method in Section \ref{app:subsec_strm_vort}.
The final contribution of mode $i$ is the outer Poisson bracket, computed as $\{\theta_i, w_i\}$. The total regularization term is then accumulated and treated as a source term in the vorticity advection equation.

In our implementation, this term is evaluated explicitly within each substage of the Runge--Kutta method. Consequently, this approach is computationally more intensive due to the elliptic solve required for each of the modes $\theta_i$.

\subsection{Time integration}\label{app:subsec_time_integration}
Time integration is carried out using the strong stability preserving third-order Runge--Kutta (SSPRK3) method \cite{gottlieb2005high}. 
Its stochastic version (StochSSPRK3) is suited for integration of dynamical systems subject to Stratonovich noise \cite{woodfield2024strong}. We highlight the stochastic method which, when removing stochastic terms, reduces to the deterministic method.

The StochSSPRK3 algorithm can be expressed as a weighted sequence of nested Euler--Maruyama steps. 
For a semi-discretized SDE with drift $f$ and diffusions $g^i$, we denote an Euler--Maruyama step by  \begin{equation}
    \EM(q^n; f, \Delta t, g^i, \DW) := q^n + \Delta t f(q^n) + \sum_i \DW g^i(q^n).
    \label{eq:EulerMaruyama}
\end{equation}
Here, $q$ is the system configuration and the superscript $n$ denotes the $n^\mathrm{th}$ time level. In what follows, $\Delta W_n^i$ are Wiener increments that are drawn independently for each $n$ and each $i$, and are distributed with variance $\Delta t$. Equation \eqref{eq:EulerMaruyama} reduces to the deterministic forward Euler time-integration method when the stochastic terms vanish. 
Similar to its deterministic counterpart, the StochSSPRK3 method is given by the three-stage sequence\begin{equation}
    \begin{split}
        q^1 &= \EM(q^n; f, \Delta t, g^i, \DW), \\
        q^2 &= \frac{3}{4}q^n + \frac{1}{4}\EM(q^1; f, \Delta t, g^i, \DW), \\
        q^{n+1} &= \frac{1}{3}q^n + \frac{2}{3}\EM(q^2; f, \Delta t, g^i, \DW),
    \end{split}
\end{equation}
where $q^1$ and $q^2$ are the internal substages and $q^{n+1}$ denotes the numerical approximation at the ensuing time instance.
In this general form, the deterministic Euler equations are integrated by taking a discretization of $\{\Delta^{-1}q, q\}$ as $f(q)$.
For simulations employing SALT, a discretization of $ \{\zeta_i, q\}$ is added as $g^i(q)$, whereas SFLT requires a discretization of $ \{q, \theta_i\}$.

In the stochastic LA SALT method, the vorticity is advected by the ensemble mean advecting velocity. Therefore, during the StochSSPRK3 algorithm, the ensemble mean is computed after each internal substage and used in the ensuing substage for each ensemble member.

\section{Quantities of interest}\label{app:qois}
Various quantities of interest are used throughout the comparison of the stochastic forcings.
Firstly, we define the integral quantities energy $E$, enstrophy $Z$, and palinstrophy $P$, as
\begin{align}
    E(\omega) &= \frac{1}{2}\int_\Omega\! |\gradperp\psi|^2\,\td\Omega = -\frac{1}{2}\int_\Omega\, \psi\omega\,\td\Omega, \\
    Z(\omega) &= \frac{1}{2}\int_\Omega\! \omega^2\,\td\Omega, \\
    P(\omega) &= \frac{1}{2}\int_\Omega\! |\nabla\omega|^2\,\td\Omega. \label{eq:palinstrophy}
\end{align}
Let $\omega^+$ denote the positive part of the vorticity. The corresponding center of mass $(\bar x^+, \bar y^+)$ is defined as \begin{equation}
\begin{split}
    \bar x^+ &= \int_\Omega\!\, x\,\omega^+ \,\td\Omega \,\bigg/ \int_\Omega \!\,\omega^+\,\td\Omega, \\
    \bar y^+ &= \int_\Omega\!\, y\,\omega^+ \,\td\Omega \,\bigg/\int_\Omega \!\,\omega^+\,\td\Omega.
\end{split} \label{eq:effective_radius}
\end{equation}
The center of mass for the negative part of the vorticity is defined analogously.
Using the center of mass, one may define an \textit{effective radius} as \begin{equation}
    R^+ = \int_\Omega \left((x-\bar x^+)^2 + (y-\bar y^+)^2\right)\omega^+\,\td\Omega
\end{equation}
and similarly for $\omega^-$. This radius is small if the vorticity is highly concentrated around the center of mass.

\bibliographystyle{abbrv}
\bibliography{refs}
\end{document}